%% file: main.tex
\pdfoutput=1
\documentclass[10pt,journal,compsoc]{Class/IEEEtran}
\input{Class/packages}
\begin{document}

\input{title}


\IEEEtitleabstractindextext{%
\begin{abstract}
  \input{abstract}  
\end{abstract}

\begin{IEEEkeywords}
    Network Slicing, Mobile Edge Computing, Resource Allocation, Market Model, Game Theory, Optimization
\end{IEEEkeywords}
}

\maketitle
\IEEEdisplaynontitleabstractindextext
\IEEEpeerreviewmaketitle

\newtheorem{theorem}{Theorem}
\newtheorem{remark}{Remark}
\newtheorem{proposition}{Proposition}

\section{Introduction}\label{sec:introduction}
\input{Content/Introduction/introduction}
\section{Related Works}\label{sec:related_works}
\input{Content/Related Works/related_works}

\section{System Model}\label{sec:sys_model}
\input{Content/System Model/system_model}

\section{Allocation Approaches}\label{sec:alloc_approaches}
\input{Content/Allocation Approaches/allocation_approaches}

\section{Market Model Properties}\label{sec:market_mod_props}
\input{Content/Market Model Properties/market_model_properties}

\section{Numerical Results}\label{sec:results}
\input{Content/Results/results}
\section{Conclusion}\label{sec:conclusion}
\input{Content/Conclusion/conclusion}
\vfill
%


%

\appendix
\label{sec:appendix}
\input{Content/Appendices/appendix}

\bibliographystyle{IEEEtran}
\bibliography{biblio.bib}

\input{Content/Authors Bio/authors_bio}

\end{document}

%% file: Class/packages.tex
\usepackage{amssymb}
\usepackage{xcolor}
\usepackage{longtable}
\usepackage[short]{optidef}
\usepackage{enumitem}
\usepackage{comment}
\usepackage{amsthm}
\usepackage{subfig}

\ifCLASSOPTIONcompsoc
  \usepackage[nocompress]{cite}
\else
  \usepackage{cite}
\fi
\ifCLASSINFOpdf
  \usepackage[pdftex]{graphicx}
\else
\fi
\usepackage{amsmath}

%% file: title.tex
%
\title{Joint Management of Compute and Radio Resources in Mobile Edge Computing: a Market Equilibrium Approach}
%
%
%
%
\author{
    Eugenio~Moro,~\IEEEmembership{Student Member,~IEEE,}
    and~Ilario~Filippini,~\IEEEmembership{Senior Member,~IEEE.}
    
    \IEEEcompsocitemizethanks{
        \IEEEcompsocthanksitem E.Moro and I.Filippini are with the Department of Electronics, Information and Bioengineering, Politecnico Di Milano, 20133 Milan, Italy.\protect\\
        E-mail: \{eugenio.moro, ilario.filippini\}@polimi.it
        }%
    \thanks{Manuscript received ****; revised ****.}%
}

%
%

%% file: abstract.tex
Edge computing has been recently introduced as a way to bring computational capabilities closer to end users of modern network-based services, in order to support existent and future delay-sensitive applications by effectively addressing the high propagation delay issue that affects cloud computing. However, the problem of efficiently and fairly manage the system resources presents particular challenges due to the limited capacity of both edge nodes and wireless access networks, as well as the heterogeneity of resources and services' requirements. To this end, we propose a techno-economic market where service providers act as buyers, securing both radio and computing resources for the execution of their associated end users' jobs, while being constrained by a budget limit. We design an allocation mechanism that employs convex programming in order to find the unique market equilibrium point that maximizes fairness, while making sure that all buyers receive their preferred resource bundle. Additionally, we derive theoretical properties that confirm how the market equilibrium approach strikes a balance between fairness and efficiency. We also propose alternative allocation mechanisms and give a comparison with the market-based mechanism. Finally, we conduct simulations in order to numerically analyze and compare the performance of the mechanisms and confirm the theoretical properties of the market model.

%% file: Content/Introduction/introduction.tex
\IEEEPARstart{T}{he} recent emergence and affirmation of Cloud Computing (CC) has determined a centralization of computing power into large data centers, providing enormous computing and storage capabilities to support the explosive growth of PaaS and IaaS-based internet services, such as on-demand streaming, computational offloading and cloud storage.\newline
As this trend is expected to continue in the future, CC will play an increasingly prominent role in web-based services by providing supercomputing-like capabilities. However, the often long distances between data centers and end users render CC unsuitable for supporting new time-sensitive use cases, such as those enabled by the fifth generation of mobile radio networks (5G)~\cite{parkvall2017}.\newline
Recently introduced Mobile Edge Computing (MEC) promises to surpass this intrinsic limit of CC by shifting the computational power toward the edge of the network, in the proximity of the Radio Access Network (RAN), in order to cut down on propagation delays and support a wide range of time-sensitive applications, such as tactile internet, agumented reality and IoT.~\cite{Abbas2018}.\newline
Additionally, MEC is expected to alleviate the burden that future data-intensive applications, such as autonomous driving and massive-IoT, might have on the networking infrastructure, as local computations avoid sending large data volumes to the cloud~\cite{shi2016}.\newline
\par
MEC is currently undergoing a developing phase and, among the different open challenges that it poses to the research community, resources allocation and management represents a particularly interesting one.\newline
Differently from what is usual in resource management studies for CC, assuming infinite network capacity between the computing nodes and the offloading source is arguably less than ideal. Indeed, Edge Nodes (EN) are expected to be deployed in close proximity to wireless access points with limited capacity. Furthermore, given the relative high density of these deployments with respect to traditional data centers, ENs can but offer limited computing capabilities in order to keep costs down and retain economical scalability of MEC. Therefore, we argue that an effective management framework cannot refrain from jointly considering both radio and computational resources.\newline
\par
The recent network and computing virtualization paradigms enable the sharing of these infrastructures among network slices offering services with different and possibly contrasting requirements, both in the computation (i.e. CPU cycles, RAM and storage) and the network (i.e. bandwidth) domain, and the possibility for  prioritization must be allowed. This service requirements diversity strongly calls for a joint optimization of both domain's resources, as uncoupling the two might lead to situations in which resources in one domain are over-provisioned with respect to the capacity of the other. Therefore, acting as a bottleneck, the latter domain drastically reduces the overall resource utilization efficiency.\newline
To add to the complexity of this challenge, MEC resources present a high degree of heterogeneity as ENs might show different configurations,  leading to diverse capacity profiles and the abundance of resources - or lack thereof - might impact the system performance in different ways, depending on the considered domain.\newline
\par
The scenario previously described requires smart resource allocation solutions. Static resource partitioning approaches, such as the one considered by 3GPP~\cite{3gppRANSHARING}, are not suitable. Indeed, the intrinsic spatial and temporal heterogeneity of both demands and resource availability suggests that elastic allocations might yield higher efficiency and better performance. Furthermore, network slices are designed to offer customized network services to tenants representing business entities whose economical performance indicators are likely to couple with the experienced service performance. 
The consequence is that tenants should be allowed to directly partake in the allocation decision process and autonomously manage their resources, in order to better match their desired techno-economic performance. Indeed, a third-party resource allocation that penalizes some tenants in favor of an overall global welfare would result in a weak, and likely unrealistic, solution.\newline
\par
Considering the environment sketched in the previous paragraphs, the following question arises: \textit{given a MEC/RAN deployment consisting of a set of heterogeneous network cells and ENs, how to efficiently allocate computing and network resources to competing services with different requirements, while guaranteeing fairness, service prioritization, and economic convenience?}\newline
\par
In light of these considerations, we propose a resource management model based on a market of resources, where each service provider is allowed to buy allocations constrained to a budget.\newline
The properties of the solution are analyzed through Game Theory and General Equilibrium Theory and the market itself is based on a instance of Fisher Market~\cite{vazirani_2007_fisher}, a well known model that has been extensively used in literature for either networking resources allocation in slicing contexts or MEC resource management, but, so far, never when the two domains are combined, to the best of the authors' knowledge. Indeed, a non trivial extension toward a joint resource management needs careful considerations on how demand and availability of each different type of resource plays its role in the overall system performance.\newline
In our proposed market model, domains' resources are dynamically priced according to availability and instantaneous demand, while buyers selfishly decide which particular bundle to buy in order to maximize their private utility function, which coincides with the number of successfully executed jobs that the allocation resulting from the market equilibrium allows for. Finally, service prioritization is enforced through budget differentiation, whose values can represent real money or, more generally, power relationships among service providers.\newline
\par
The remainder of this article is structured as follows. Section~\ref{sec:related_works} presents some relevant related works and point out the novel contribution of this work with respect to the others in literature. Section~\ref{sec:sys_model} presents the system model formulation and Section~\ref{sec:alloc_approaches} presents the proposed resource allocation approaches, whose properties are theoretically analyzed in Section~\ref{sec:market_mod_props}. Finally, simulation results are shown in Section~\ref{sec:results}. The article concludes with Section~\ref{sec:conclusion} that includes final remarks.

%% file: Content/Related Works/related_works.tex
\begin{figure}[!t]
\centering
\includegraphics[width=3.5in]{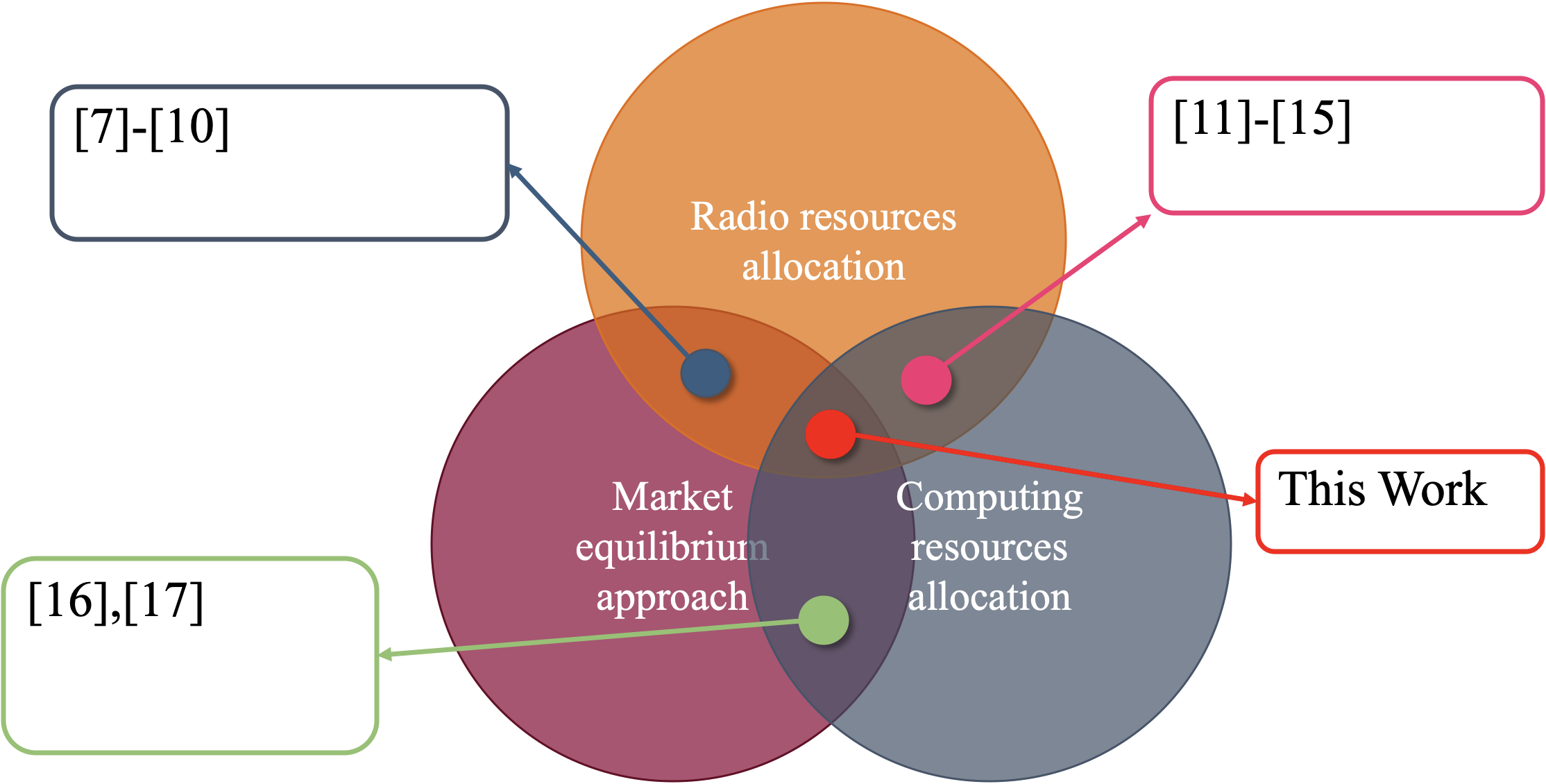}
\caption{Literature study areas and their intersections}
\label{fig:soa_intersection}
\end{figure}
The main results of this work is based on game theory, which has been extensively utilized in network resource management~\cite{lasaulce2011game}. In the context of game theory-based network resources management for slicing, authors in~\cite{journalStoccarda} propose a single-cell budget-free market with prices computed according to some function of the cell load and in~\cite{jiang2017} an auction model for the resource allocation problem is proposed. Authors of~\cite{Caballero2017} propose a budget-constrained market model (extended to admission control in~\cite{caballero2018}) that , only from the radio resources management point of view, faces the problem following a rationale similar to the one of our proposal. \newline
\par
In the multi-server computing resources management literature, the multi-resource allocation problem has been first addressed in~\cite{Wang2014}, where the well known Dominant Resource Fairness concept~\cite{Ghodsi2011} has been extended to multiple heterogeneous servers scenarios. An external resource, generically identified as bandwidth, was later introduced in~\cite{Meskar2018} as an extension to the previous approach. More refined approaches are to be found in~\cite{Xiang2020}, where the problem of the joint resource optimization to provide the desired QoS to all mobile users and traffic types is formulated as a mixed-integer nonlinear program, and in~\cite{Sardellitti2015}, where the joint optimization is aimed at minimizing users' energy consumption while meeting latency constraints.\newline
Both~\cite{Nguyen2018} and~\cite{Nguyen2019} approach the computing resource allocation problem employing game-theoretical tools, proposing a Fisher Market-based model for the management of MEC resources, however the first work studies the system only from a high-level perspective at the EC platform (i.e. not considering the radio access segment) and the second avoids managing radio resources claiming transmission delay to be constant and negligible. We argue this assumption to be less than ideal, as future services are expected to challenge the radio access segment of the network, either by large job payloads (i.e. video trans-coding, image processing), large number of simultaneous requests (i.e. massive-IoT) or a combination of both (i.e. autonomous driving and connected cars) and thus network resource management becomes relevant~\cite{chiang2016}.\newline
\par
In the previous analysis, 3 main areas of study can be identified: radio resources allocation, computing resources allocation and market equilibrium-based allocation approaches, as pictured in Figure~\ref{fig:soa_intersection}. While the related works we presented cover up to two of the aforementioned areas, we were not able to find any work that satisfyingly cover all 3 endeavours, thus we claim this paper to be the first covering the problem of allocating resources both in the radio and computing domains in the context of network slicing services and MEC by means of a techno-economic market of resources.\newline
\par
The main contribution of this work can be further summarized as follows:
\begin{itemize}
    \item We formulate a system model for the aforementioned environment that comprises for the first time in literature the joint management of resources in both MEC and RAN domains.
    \item We propose a novel market-based allocation mechanism and give a convex program formulation to find a ME. Additionally, we propose some alternative allocation mechanisms.
    \item We formally analyze the theoretical properties of the proposed mechanisms, with a focus on the efficiency and fairness of the ME approach.
    \item We give an extensive numerical evaluation of the proposed mechanism through simulations.
\end{itemize}

%% file: Content/System Model/system_model.tex
\begin{figure}[!t]
\centering
\includegraphics[width=0.4\textwidth]{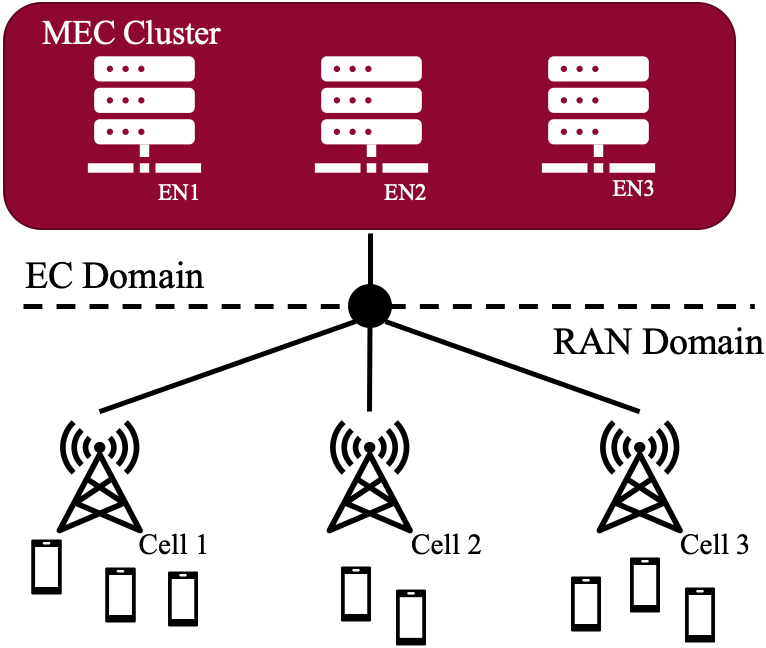}
\caption{High level system architecture.}
\label{fig:sys_archi}
\end{figure}

Consider a system in which mobile end users run jobs on a shared cluster of heterogeneous edge nodes. Users can connect and upload job payloads to the EN cluster via a shared RAN composed of heterogeneous network cells. Figure~\ref{fig:sys_archi} shows an high level graphical representation of this generic architecture, where we also highlight the two different resource domains, MEC and RAN respectively.\newline
Our proposal aims to jointly manage the resources of the system, by taking into account the diversity of demand and availability of resources in each domain, as well as the different impact that each resource type might have on the system performances.\newline 
In the MEC domain, the resources to be managed are those typical of a cloud computing system, namely CPU, RAM and storage. Let $\mathcal{M}$ be the set of $M$ edge nodes in the MEC cluster and $\mathcal{R}$ the set of $R$ resource types that is offered by the nodes, then by $D_{m,r}^\text{MEC}$ we express the available capacity of resource type $r \in \mathcal{R}$ in node $m \in \mathcal{M}$. Since we consider an heterogeneous MEC cluster, $D_{m,r}^\text{MEC}$ can in principle be different for each resource and node, in order to model real life scenarios in which ENs have different installed capacity, as well as those situations in which a fraction of the node capacity is reserved for external purposes and thus not available for scheduling. Finally, $D_{m,r}^\text{MEC}$ can be represented in the unit of measure that best suits the resource type, i.e. number of CPUs or RAM size.\newline
Differently, in the RAN domain we aim at optimizing the radio resources allocation in each BSs to deliver the network throughput that end users need in order to upload their job payloads to the MEC cluster. Let $\mathcal{C}$ be the set of $C$ cells that can be used to access the MEC cluster, we then define the available capacity\footnote{This generic definition of network capacity can be adapted to a series of different radio access technologies. For instance, $D^\text{RAN}_c$ can be expressed in terms of available physical resource blocks for 4G and 5G networks, time slots for TDM systems and available spectrum for FDM systems.} of each cell $c \in \mathcal{C}$ as $D^\text{RAN}_c$. For the sake of simplicity, in the remainder of this paper we consider resources to be expressed in terms of cell spectrum portions and we suppose the presence of a slicing-aware MAC scheduler capable of dynamically enforcing these constraints, such as the one proposed in~\cite{mandelli2019}. We must note that, since RAN access capacity is limited, we assume it to be the bottleneck of the connection to the MEC cluster, therefore we do not impose any limitation on the backhaul network capacity that connects each cell with any of the ENs in the MEC cluster. Furthermore, once the job payload enters the MEC domain, it can be freely scheduled for execution in any node of the cluster. 
\subsection{Demand-based Service Characterization}
We consider each end user to be associated with a service provider (SP), who owns the network slice and offers its network and computational resources as one specific service to the subscribed users.\newline
The specific service offered by a SP can be characterized in terms of computing and networking resources needed for the completion of associated jobs. For instance, there might be a CPU-intensive service whose CPU resource demand could be relatively higher than its memory requirement, or a network-intensive type of service whose job's payload is larger than others' payloads, thus requiring higher bandwidth allocation.\newline
Consequently, it is likely that jobs originating from end users subscribed to the same SP present similar resource requirements and thus we define a set of demand requirements characterizing each service provider and we call it demand profile\footnote{It must be noted that, while we do not exclude the presence of two or more service providers with the same demand profile, we consider each SP to offer only one type of service and thus to be associated with only one demand profile.}.\newline
Let $\mathcal{S}$ be the set of $S$ service providers, then we define $d_{s,r}^\text{MEC}$ as the minimum amount of MEC resource of type $r \in \mathcal{R}$ needed to complete one job of service provider $s \in \mathcal{S}$ in a timely fashion. Similarly, $d_{s,c}^\text{RAN}$ indicates the minimum amount of network resources needed to successfully upload one job of service provider $s \in \mathcal{S}$ through cell $c \in \mathcal{C}$. We must note that, while $d_{s,r}^\text{MEC}$ does not depend on the specific EN where the job is executed, $d_{s,c}^\text{RAN}$ must depend on the cell where the payload is uploaded, as different channel quality might require higher or lower bandwidth requirements for the same payload size\footnote{In case end users of a given service provider experience strong differences in channel quality in a single cell, then it is always possible to divide end users subscribed to a single service into fictitious sub-services that differ from the original only in the network resource requirements.}.
\newline
\par
Given the proximity between end users and MEC clusters, delay-sensitive services are often considered as the main application of edge computing, while delay-tolerant services are better suited to be executed in centralized cloud deployments~\cite{Mach2017}. For this reason, we consider all jobs of service providers in $\mathcal{S}$ to show a certain delay-sensitiveness and this must be taken into account when defining MEC resource demands $d_{s,r}^\text{MEC}$. This quantity expresses the minimum amount of resources of each type that, if reserved to service $s$, allows for one job to be completed in a time interval that is acceptable from a delay-sensitiveness standpoint. Indeed, the completion time of a computing job is strictly related to the amount of computing resources reserved: for instance a higher CPU-time share leads to a faster computation, as well as a larger RAM allocation leads to fewer disk accesses and thus an execution speed up. In this work we consider $d_{s,r}^\text{MEC}$ as given, supplied by service providers, and we suppose that, if $d_{s,r}^\text{MEC}$ or more resources are allocated for each resource type, then one job can be considered as successfully executed in due time. Moreover, we allow for numerous jobs to be executed simultaneously if multiples of this quantity are allocated.\newline
In the RAN domain, similar considerations can be done when computing the amount of required network resources $d_s^{RAN}$. Specifically, both the transmission time (i.e. payload upload time for any given allocation) and the technological delay (i.e. as low as $1ms$ for 5G~\cite{Parvez2018}) can be taken into account when defining the smallest amount of network resources needed for a successfully job payload transmission in due time. Same as before, we consider $d_s^{RAN}$ as given and we allow the number of timely uploaded payloads to scale with larger allocations.\newline
%
\subsection{Bottlenecks Among Resources and Domains}
In multi-resource allocation scenarios, the heterogeneous availability of resources and the demand diversity across different users of the shared infrastructure poses some non-trivial challenges to the design of an efficient and fair allocation mechanism~\cite{Meskar2018}. One of the aspects that must be not overlooked is the presence of resource bottlenecks and the emergence of a dominant resource~\cite{Ghodsi2011}.\newline 
Considering the MEC domain of the proposed architecture, a specific service provider can experience a resource bottleneck in a particular edge node of the cluster when its associated demand profile is such that there exists a resource type demand that exhausts the availability in said node. We call this resource \textit{dominant} and when the node capacity for such resource is depleted, then no more simultaneous jobs can be scheduled, even if other resource types are available in the EN. This bottleneck issue cannot be avoided and it is further exacerbated by the heterogeneity of both demand profiles and server capacity configurations.\newline
On the other hand, there is no risk of resource bottleneck in the RAN domain, since the allocable resource is of one type only. However, given the time sensitivity of jobs, we do not allow for queuing at the domains interface, implying that the total system throughput is limited by the least performing domain. This can be considered as a bottleneck as well, since no further jobs can be accommodated in the system once the capacity of one domain reaches its saturation point.
\newline In our proposed model, we mathematically capture the presence of these resource bottlenecks in the system by a proper definition of the service provider utility function.
\subsection{Service Utility Function}
Once the allocation decision is carried out, resources are reserved for each provider and no variation is expected before the subsequent iteration.\newline
At this point, service providers are free to utilize the assigned capacity according to any scheduling and admission control policy they see fit. However, there exists a theoretical maximum number of jobs that can be simultaneously executed for each service provider after which a per-job allocation of $d_{s,r}^\text{MEC}$ and $d_s^{RAN}$ cannot be guaranteed anymore. Consequently, only a limited number of jobs can be concurrently accommodated by the system at any point of time, if the delay requirements are to be met. By expressing this theoretical limit in form of a service utility function, we give a mean to numerically evaluate this maximum number of simultaneously executable jobs.\newline
Note that this approach is perfectly in line with realistic virtualization scenarios in which network and cloud operators rent pure capacity to service providers in terms of job processing resources and are agnostic about how they will use such a capacity to serve its customer job load.\newline
\par
Let $x_{s,m,r}$ be the amount of computing resource type $r$ reserved to service provider $s$ in node $m$. Then for any resource $r$, $\frac{x_{s,m,r}}{d_{s,r}^\text{MEC}}$ represents the maximum number of concurrent jobs of $s$ that such an allocation allows to be executed in EN $m$. However, the actual number of concurrently executed jobs is limited by the dominant resource, thus we must consider the minimum of this quantities over all the computing resource types. By summing over the nodes in the system, we get the maximum number of jobs that can be simultaneously executed with acceptable per-job performance in the MEC domain for service provider $s$:
\begin{equation}
\label{eq:mec_tp}
    J_s^\text{MEC}\left(\textbf{X}_s\right) = \sum_{m \in \mathcal{M}} \min_{r \in \mathcal{R}}\left\{ \frac{x_{s,m,r}}{d_{s,r}^\text{MEC}}  \right\},
\end{equation}
where $\mathbf{X}_s=\left( x_{s,m,r}\right)\in \mathbb{R}^{M\times R}$ is the computing resources allocation matrix.\newline
Similarly, by letting $y_{s,c}$ be the network resources allocated in cell $c$ for service provider $s$, the ratio between $y_{s,c}$ and $d_{s,c}^\text{RAN}$ can be identified as the largest number of job payloads of service $s$ that can be simultaneously sent to the MEC cluster through cell $c$. Summing over all the cells in the system, we get the number of payloads of service provider $s$ that can traverse the RAN domain for a given network resources allocation vector $\mathbf{Y}_s=\left\{ y_{s,1},...,y_{s,C}\right\}$:
\begin{equation}
\label{eq:ran_tp}
    J_s^\text{RAN}\left(\mathbf{Y}_s\right)=\sum_{c \in \mathcal{C}}\frac{y_{s,c}}{d_{s,c}^\text{RAN}}
\end{equation}
Finally, being the number of concurrently executed jobs limited by the least performing domain, we can express the service provider utility as follows.
\begin{equation}
\label{eq:serv_tp}
    u_s\left(\mathbf{X}_s,\mathbf{Y}_s\right) = \min\left\{ J_s^\text{MEC}\left(\mathbf{X}_s\right), J_s^\text{RAN}\left(\mathbf{Y}_s\right)\right\}.
\end{equation}

%% file: Content/Allocation Approaches/allocation_approaches.tex
In this section we present different approaches to the allocation problem in the settings described in Section~\ref{sec:sys_model}, whose notation is summarized in Table~\ref{tab:notation}.\newline 
The main focus of this work resides in the definition of a techno-economic market model where service providers individually buy resources in each edge node and cell, which will be then dedicated to the execution of jobs originating from their end users.\newline
The equilibrium points of such market model coincide with the outcome of the proposed market-based allocation mechanism and we formulate a convex optimization program in order to numerically compute equilibria.\newline 
We also present other possible allocation approaches, such as a pure optimization of the allocation decisions and a more traditional proportional sharing scheme. These will be used as baselines for the market solution in section~\ref{sec:results}, where we compare the features of the different approaches and their effectiveness in terms of system performance fairness from a numerical point of view.
\begin{table}[t]
\centering
\caption{Notations}
\label{tab:notation}
\input{Content/Allocation Approaches/notation_table}
\end{table}
\subsection{Market Equilibrium Approach}
We propose a techno-economic market of resources model $\Gamma$ based on a Fisher Market~\cite{vazirani_2007_fisher} model. In $\Gamma$, buyers are represented by the individual service providers in set $\mathcal{S}$. The goods in the market are represented by each resource type in the system, therefore we identify $M\times R + C$ different divisible goods.\newline
The market capacity of each good is given by the availability of each resource, namely~$D_{m,r}^\text{MEC}$ and~$D_c^\text{RAN}$. Each resource has an associated unitary price $p_{m,r}^\text{MEC}$ (for MEC domain resources) and $p_c^\text{RAN}$ (for RAN domain resources) that SPs must pay if they intend to acquire it. Resources bought by providers in the market constitute their allocation vectors~$\mathbf{X}_s$ and~$\mathbf{Y}_s$, yielding performances that each SP $s$ can privately evaluate by means of the utility function $u_s$, as defined in Eq.(\ref{eq:serv_tp}).\newline
In this framework, we expect service providers to act as rational agents, all having the goal of pursuing their own interest by buying resource bundles that maximize their utility (i.e. maximizing the number of simultaneously executable jobs), while being constrained by their budget limit $B_s$.\newline
As it will be shown later, resource pricing in the proposed approach follows the natural law of demand and supply by assigning comparatively higher prices to those resources who are more desired and less available. Accordingly, prices are expected to have a load balancing effect on the allocations, as SPs are incentivized to buy resources where congestion is lower.\newline
Since both network and computing resources are allocated according to the rational decisions of providers and their interactions, the analysis of the resulting allocation mechanism is carried out utilizing tools from game theory~\cite{Skyrms1997} and general equilibrium theory~\cite{Colell1995Book}, focusing on the characterization of the market equilibrium (ME) outcome for this market model.\newline
Resource bundles $\left(\mathbf{X}_1,\dots,\mathbf{X}_S,\mathbf{Y}_1,\dots,\mathbf{Y}^*_S\right)$ and corresponding prices $\left( p_{m,r}^\text{MEC}, p_{c}^\text{RAN} \right)$ are said to induce a market equilibrium if the following conditions are true~\cite{branzei2014}:
\begin{enumerate}[label=(\roman*)]
    \item \label{cond:me1}\textbf{Optimal Goods}: Every service provider buys only those goods that yield the maximum utility per unit of money, also known as maximum \textit{bang-per-buck};
    \item \label{cond:me2}\textbf{Market Clearing}: Either resources are fully allocated, or the corresponding price is zero. Furthermore, each service provider exhausts its budget.
\end{enumerate}
The first condition is intended as a satisfaction of the rationality and selfishness of the buyers, whose unique interest is to maximize the return of their market investment in the manner prescribed by their utility functions; intuitively, no equilibrium could be otherwise established.\newline
Condition~\ref{cond:me2} states that the offer of each resource can either meet the demand and thus be priced, or be in a surplus condition and given away for free. Additionally, stability can be reached only when each buyer spends its entire budget, entailing that any unused fraction can be indeed spent to increase the utility. \newline 
The market equilibrium approach we propose lets the equilibrium-inducing resource bundles coincide with the actual allocation outcome, thus exploiting the previously mentioned efficiency and stability properties of equilibria.\newline
In other words, after the market parameters are defined and the equilibrium is reached by the buyers, the resources bought by each service provider are then reserved in the system.\newline
\par
We now present a convex optimization program whose solution is proven in Section~\ref{sec:market_mod_props} to be in the set of market equilibria of $\Gamma$, meaning that it can be effectively used as an algorithm to find the market equilibrium and thus compute the resulting allocation in the market equilibrium approach.\newline
Let variables $x_{s,m,r},\, y_{s,c},\, u_s$ and sets $\mathcal{S},\, \mathcal{M},\, \mathcal{R}, \mathcal{C}$ of the program be defined as done in Sec.~\ref{sec:sys_model} and let $j_{s,m}$ be the variable associated with the number of jobs of SP $s$ executed in EN $m$.
\begin{maxi!}[3]
  {\mathcal{X},\mathcal{Y}}{\sum_{s \in \mathcal{S}}B_s \log(u_s)}{}{}\label{opt:obj}
  \addConstraint{j_{s,m}}{\leq \frac{x_{s,m,r}}{d_{s,r}^\text{MEC}},\qquad}{\forall s \in \mathcal{S}, m \in \mathcal{M}, r \in \mathcal{R}}\label{opt:res_bottleneck}
  \addConstraint{u_s}{\leq \sum_{m \in \mathcal{M}}j_{s,m},\qquad}{\forall s \in \mathcal{S}}\label{opt:mec_bottleneck}
  \addConstraint{u_s}{\leq \sum_{c \in \mathcal{C}}\frac{y_{s,c}}{d_{s,c}^\text{RAN}},\qquad}{\forall s \in \mathcal{S}}\label{opt:bw_bottleneck}
  \addConstraint{\sum_{s \in \mathcal{S}}x_{s,m,r}}{\leq D_{m,r}^\text{MEC}, \qquad}{\forall m \in \mathcal{M}, r \in \mathcal{R}}\label{opt:mec_cap}
  \addConstraint{\sum_{s \in \mathcal{S}}y_{s,c}}{\leq D_{c}^\text{RAN}, \qquad}{\forall c \in \mathcal{C}}\label{opt:bw_cap}
  \addConstraint{x_{s,m,r},\,y_{s,c},\,j_{s,m}}{\geq 0, \qquad}{\forall s \in \mathcal{S}, m \in \mathcal{M}, r \in \mathcal{R}, c \in \mathcal{C}}\label{opt:positive_vars}
\end{maxi!}
\newcommand{\referenceEG}{\ref{opt:obj}-\ref{opt:positive_vars}}
Constraints~(\ref{opt:res_bottleneck}) to~(\ref{opt:bw_bottleneck}) linearize the utility function defined in Eq.~(\ref{eq:serv_tp}). In more details, constraint~(\ref{opt:res_bottleneck}) expresses the $\min$ operator in Eq.~(\ref{eq:mec_tp}), while constraint~(\ref{opt:mec_bottleneck}) forces the utility not to be larger than the number of jobs that can be concurrently executed in the MEC domain. Similarly, constraint~(\ref{opt:bw_bottleneck}) limits the utility $u_s$ to the maximum number of payloads that can be simultaneously processed by the RAN domain, as defined in Eq.~(\ref{eq:ran_tp}).\newline
Constraint~(\ref{opt:mec_cap}) is such that the total allocation for each resource in each EN does not exceed the node capacity. In similar fashion, constraint~(\ref{opt:mec_cap}) limits the allocated network resource in each cell to the available capacity.\newline
The solution of such program is represented by an optimal utility vector $\mathbf{u}^*=\left(u_1^*,\dots,u_S^*\right)$ that gives the result of the allocation mechanism by measuring how many jobs coming from end users of each service provider can be concurrently accommodated in the system, together with the allocations that produce such performances, namely $\mathbf{X}^*=\left( \mathbf{X}_1^*,\dots,\mathbf{X}_S^*\right)$ and $\mathbf{Y}^*=\left( \mathbf{Y}_1^*,\dots,\mathbf{Y}_S^*\right)$.\newline
Optimal prices $p_{m,r}^{*\text{MEC}}$ and $p_{c}^{*\text{RAN}}$ do not directly appear in~(\ref{opt:obj}-\ref{opt:positive_vars}) but, as strong duality holds, they can be extracted as dual variables~\cite{Horst1984} associated with capacity constraints~(\ref{opt:mec_cap}) and~(\ref{opt:bw_cap}) and, as such, represent the rate of change of the objective function associated with any change in the capacity constraint. In other words, those resources whose increment or decrement of capacity would impact the objective the most have the highest prices, suggesting a direct proportionality between prices and resource demand. Additionally, prices are such that the optimal resource bundle of each provider does not present costs higher than the provider's budget.\newline
Finally, the objective function in Eq.~(\ref{opt:obj}), which will be exhaustively analyzed in later sections, combines the individual provider's utility functions in such a way that a balance between efficiency and fairness is promoted.

\subsection{Social Optimum Approach}
We propose an approach based on linear optimization that aims to allocate resources such that the overall number of jobs that can be simultaneously executed without incurring in performance degradation is maximized.\newline
While this mechanism operates on exactly the same system model as of market model $\Gamma$, its resulting allocation cannot be guaranteed to represent a market equilibrium of any kind. Consequently, this mechanism does not consider service providers as entities capable of making decisions, but simply manages the system resources in order to maximize the performances according to some objective functions.\newline
On this matter, we propose two different objective functions to maximize system performances: a pure maximization of the sum of utilities (i.e. total executed jobs) and a sum of utilities weighted by each provider's budget.
The first objective function, also known as Social Optimum (SO), is defined as follows:
\begin{equation}
    \sum_{s \in \mathcal{S}} u_s,
\end{equation}
This objective leads to a maximization of system performance regardless of the budget difference among services and it effectively gives an indication of the maximum number of jobs that the system can simultaneously process at any moment.\newline
The second objective function is instead defined as:
\begin{equation}
    \sum_{s \in \mathcal{S}} B_su_s
\end{equation}
and includes a certain degree of fairness in the solution by giving more weight to service providers with larger budgets, replicating the advantage that such providers have in the market model $\Gamma$. This type of solution is also known as Weigthed Social Optimum (WSO).\newline
Finally, the optimization program employed for this approach shares exactly the same constraints as the previously formulated optimization problem~(\ref{opt:obj}-\ref{opt:positive_vars}), regardless of the objective function choice.
\subsection{Proportional Sharing Scheme}
This resource sharing scheme was proposed as a static mean of allocating resources in network slices~\cite{Costa2013} and has been used as a baseline allocation scheme both in resource management for network slicing~\cite{Caballero2017} and mobile edge computing~\cite{Nguyen2019}.\newline
Proportional sharing (PS) allocation mechanism is such that each service provider receives an amount of the capacity of each resource in the system that is proportional to its budget and it can be seen as an extension of the well known \textit{generalized processor sharing}~\cite{parekh1993} algorithm to a multi-resource scenario.\newline
Formally, let $\hat{x}_{s,m}^r$ and $\hat{y}_{s,c}$ be respectively the computing and network resources allocations of service provider $s$ resulting from proportional sharing, then:
\begin{flalign}
    &\hat{x}_{s,m,r} = \frac{B_s}{\sum_{s' \in \mathcal{S}}B_{s'}}D_{m,r}^\text{MEC}, &\forall s \in \mathcal{S}, m \in \mathcal{M}, r \in \mathcal{R},\\
    &\hat{y}_{s,c} = \frac{B_s}{\sum_{s' \in \mathcal{S}}B_{s'}}D_c^\text{RAN}, &\forall s \in \mathcal{S}, c \in \mathcal{C}.
\end{flalign}
Once these quantities are computed, proportional sharing utilities $\hat{u}_s$ can be computed by means of Eq.~(\ref{eq:serv_tp}).\newline
Clearly, this allocation scheme does not take into account the single provider's demands, nor the load conditions of ENs and cells. Indeed, it simply allocates all the available resources such that service providers with larger budgets get comparatively larger shares. For this reason, it is considered as a static allocation scheme, as it does not reflects any change in the system conditions. 

%% file: Content/Allocation Approaches/notation_table.tex
\begin{tabular}{|l|l|} 
\hline
\textbf{Notation}                               & \textbf{Meaning}                                             \\ 
\hline
$\mathcal{S},S$                                 & SP set and set cardinality                              \\ 
\hline
$\mathcal{M},M$                                 & EN set and cardinality                                       \\ 
\hline
$\mathcal{R},R$                                 & Computing resource type set and cardinality                  \\ 
\hline
$\mathcal{C},C$                                 & Cell set and cardinality                                     \\ 
\hline
$D_{m,r}^\text{MEC}$                                         & Resource $r$~capacity in EN $m$                              \\ 
\hline
$D_c^\text{RAN}$                                     & Network resource capacity in cell~$c$                        \\ 
\hline
$d_{s,r}^\text{MEC}$                                         & Resource $r$~needed for SP $s$ job completion~          \\ 
\hline
$d_{s,c}^\text{RAN}$                                 & Resource needed in cell $c$ for SP $s$ payload upload   \\ 
\hline
$x_{s,m,r}$                                     & Resource~$r$~allocated to SP $s$~in EN $m$              \\ 
\hline
$y_{s,c}$                                 & Network resource allocated to SP $s$~in cell~$c$        \\ 
\hline
$\mathbf{X}_s$                                  & Computing resource allocation matrix for SP $s$         \\ 
\hline
$\mathbf{Y}_s$                                  & Network resource allocation vector for SP~$s$           \\ 
\hline
$J_s^\text{MEC}\left(\mathbf{X}_s\right)$            & Concurrent jobs in MEC domain given allocation~$\mathbf{X}_s$  \\ 
\hline
$J_s^\text{RAN}\left( \mathbf{Y}_s \right)$          & Simultaneously uploaded jobs given allocation~$\mathbf{Y}_s$           \\ 
\hline
$u_s \left( \mathbf{X}_s, \mathbf{Y}_s \right)$ & Service provider utility                                          \\ 
\hline
$p_{m,r}^\text{MEC}$                                         & Price of computing resource $r$ in EN $m$                    \\ 
\hline
$p_{c}^\text{RAN}$                                       & Price of network resources in cell $c$                       \\ 
\hline
$B_s$                                           & Budget of SP $s$                                        \\
\hline
\end{tabular}

%% file: Content/Market Model Properties/market_model_properties.tex
This section presents a theoretical analysis of the previously defined market model $\Gamma$ and of the other proposed mechanisms.\newline
We first discuss the existence of a market equilibrium, in particular we prove that the solution of~(\referenceEG) is always a ME for $\Gamma$. Afterwards, we turn our focus on the analysis of efficiency and fairness of the mechanisms proposed in Section~\ref{sec:alloc_approaches}.\newline
\subsection{Market Equilibrium}
The existence of a market equilibrium for the Fisher market model is guaranteed in the most generalized settings under the mild condition that each resource is desired by at least one buyer and each buyer desires at least one resource~\cite{MAXFIELD199723}. In the remainder of this work, we consider this condition to always hold without loss of generality, since we assume that jobs require a non-zero amount of all types of resources in the system to be successfully executed\footnote{If this was not the case, then the non-desired resource could be excluded from the market model without affecting the allocation solution.}.\newline
Having established the existence of a ME for $\Gamma$, we turn our focus to its computation. Sperner's coloring~\cite{Scarf1967} could be used to find an equilibrium point, although with an high computational inefficiency. A more efficient and well known alternative is to employ a convex optimization formulation, called Eisenberg-Gale (EG) program, firstly proposed as a mean to find an equilibrium in case of linear utilities~\cite{Eisenberg1959} and then extended to more general functions~\cite{Eisenberg1961}.\newline
Program~(\referenceEG) is indeed a variant of the EG formulation applied to the proposed market model, where constraints (\ref{opt:res_bottleneck}),~(\ref{opt:mec_bottleneck}) and~(\ref{opt:bw_bottleneck}) are added to the usual EG formulation in order to express the particular bottleneck-defined service provider utility function that we employ in the proposed model, as defined in Eq.~(\ref{eq:serv_tp}). With the following theorem we prove that the solution of such formulation is in the set of market equilibria of $\Gamma$:
\begin{theorem}
\label{theo:me_solution}
Optimal variables $\mathbf{X}^*=\left( \mathbf{X}_1^*,\dots,\mathbf{X}_S^*\right)$ and $\mathbf{Y}^*=\left( \mathbf{Y}_1^*,\dots,\mathbf{Y}_S^*\right)$ and corresponding optimal prices of program~(\referenceEG) induce a market equilibrium for $\Gamma$.
\end{theorem}
\par
\textit{Proof:} See appendix.\newline
\par
This theorem allows~(\referenceEG) to be effectively used as a tool for computing a ME for $\Gamma$. In particular,~(\referenceEG) can be applied to any instance of $\Gamma$ and system resources can be allocated according to the optimal variables~$\mathbf{X}^*$ and~$\mathbf{Y}^*$. Resource prices can be extracted from the solution as dual variables of capacity constraints and per-provider and system performances can be readily evaluated using optimal utilities.\newline
Additionally, program~(\referenceEG) is in the class of convex maximization problems, which can be efficiently solved with modern numerical methods such as the interior-point method~\cite{potra2000}.\newline
The objective function is strictly concave and thus the problem has at most one optimal point~\cite{boyd2004convex}, meaning that the set of market equilibria utilities of $\Gamma$ obtained through~(\referenceEG) actually contains a single element, as summarized by the following remark.
\begin{remark}
\label{remark:uniqueness}
The market equilibrium utilities obtained through program~(\referenceEG) is unique.
\end{remark}
This result is not trivial, since there is no guarantee that the equilibrium of a Fisher market is unique and different algorithms might converge to different equilibria depending on the initialization, as is it typical for best response dynamics~\cite{ellison1993} or Sperner's coloring. 
\subsection{Efficiency Properties}
Generally speaking, the efficiency of an allocation mechanism quantifies the loss in overall system performances with respect to the maximum allowed by the system capacity.\newline
There is no guarantee that a market equilibrium based approach efficiently utilizes system resource and, in general, the efficiency of such solution highly depends on the parameters of the specific allocation instance. For this reason, this section presents some general efficiency results based on bounds, while a numerical evaluation of this metric is given in Section~\ref{sec:results}.\newline
We define the efficiency of an allocation mechanism as the ratio between the resulting total number of concurrently executed jobs and the maximum number of jobs that the capacity of the system allows for simultaneous execution.\newline
For any instance $\gamma$ of $\Gamma$, let $SW_{me}(\gamma)$, $SW_{so}(\gamma)$, $SW_{ps}(\gamma)$ be the social welfare (i.e. the sum of service provider utilities) of the solutions given by market equilibrium, social optimum and proportional sharing allocation mechanisms, respectively. These values are unique\footnote{This is true for market equilibrium, as stated by Remark~\ref{remark:uniqueness}, but also for social optimum, since it is the optimal value of the objective of a linear program, and for proportional sharing by definition.} for any instance of $\Gamma$ and the efficiency of an allocation mechanism can be evaluated by comparison with $SW_{so}(\gamma)$, since it represents the maximum social welfare for the given instance.\newline
In details, we define the following efficiency expressions:
\begin{align}
    \eta_{me}(\gamma) = \frac{SW_{me}(\gamma)}{SW_{so}(\gamma)},\\
    \eta_{ps}(\gamma) = \frac{SW_{ps}(\gamma)}{SW_{so}(\gamma)}.
\end{align}
We claim that the market equilibrium approach yields efficiency at least as high as proportional sharing, namely $\eta_{me}(\gamma)\geq \eta_{ps}(\gamma)$, for any market instance $\gamma$. This results comes from the following theorem:
\begin{theorem}
\label{theo:shar_inc}
For a given market instance, let $(u_1^*,\dots,u_s^*)$ be the market equilibrium utilities and $(\hat{u}_1,\dots,\hat{u}_s)$ be the utilities given by the proportional sharing mechanism. Then $u_s^*\geq \hat{u}_s$ for each service provider $s \in \mathcal{S}$.
\end{theorem}
\par
\textit{Proof.} See appendix.\newline
\par
Interestingly, this theorem goes beyond comparing the efficiency of the two allocation approaches by showing how each service provider cannot lose utility by taking part in the market mechanism with respect to the utility obtained through a proportional division of resources.\newline
This property is also known as \textit{sharing incentive} and it is widely considered as a desirable goal for multi resource allocation mechanism applied to data centers~\cite{poullie2018}, since it confirms that the players can all be better off by partaking in the market mechanism than to recur to externally imposed proportional allocations.
\newline
Next, we compare the market equilibrium efficiency with respect to the maximum efficiency, given by the social optimum solution.\newline
In particular, in~\cite{zhang2005} it is proven that proportional sharing mechanism has an efficiency asymptotically lower bounded by $1/\sqrt{S}$ under mild conditions for the utility functions holding for $\Gamma$ and the bound is tight.
\newline 
By Theorem~\ref{theo:shar_inc}, proportional sharing cannot yield efficiencies higher than market equilibrium, thus the bound also holds for $\eta_{me}$ and the result is summarized in the following remark:
\begin{remark}
\label{remark:efficiency_bound}
$\eta_{me}(\gamma)$ is asymptotically lower bounded by $1/\sqrt{S}$ for any $\gamma$ instance of $\Gamma$. 
\end{remark}
Additionally, the remark above can be employed to characterize the price of anarchy (PoA) of the market equilibrium approach, a well known measure of efficiency in equilibrium theory~\cite{koutsoupias1999}. PoA is defined as the ratio between the worst case social welfare of the equilibrium solution and the optimal social welfare obtained through optimization and it can be expressed as a function of the previously defined ME efficiency:
\begin{equation}
    PoA_{me} = \frac{1}{\min_{\gamma \in \Gamma}\eta_{me}(\gamma)},
\end{equation}
that is, the inverse of the worst case ME efficiency.\newline
Given this definition, an upper bound for $PoA_{me}$ is derived in the following remark:
\begin{remark}
$PoA_{me}$ is asymptotically upper bounded by $\sqrt{S}$.
\end{remark}
\par
The last efficiency-related trait of the proposed allocation mechanism analyzed in this section is the well known \textit{Pareto Efficiency}~(PE)~\cite{Colell1995Book}.\newline
A feasible utility vector $(u_1,\dots,u_S)$ obtained through any allocation mechanism is said to be Pareto optimal if there is no other feasible utility vector $(u_1^*,\dots,u_S^*)$ such that $u_s^*\geq u_s$ for all $s \in \mathcal{S}$ and $u_s^*> u_s$ for at least one $s$.\newline
Any allocation mechanism whose utilities are Pareto optimal is said to be PE and this property is often considered as a minimal notion of efficiency.\newline
It is easily verifiable that both social optimum and proportional fairness solutions are PE, while the same is not generally true for equilibrium points. However, under some mild conditions holding also for $\Gamma$, the \textit{First Fundamental Theorem of Welfare Economics}~\cite{Colell1995Book} states that any equilibrium of a competitive market is PE. 
\subsection{Fairness Properties}
Fairness is a highly sought-after concept when designing resource allocation mechanisms and it is unanimously considered as fundamental in guaranteeing a certain level of QoE among heterogeneous services.\newline
However, differently from efficiency, there is no univocal definition of fairness. In this work, fairness is analyzed both through a numerical evaluation of a popular fairness index and by showing that the market equilibrium satisfies some fairness properties that are commonly desired in the resource allocation literature.\newline
To numerically quantify the fairness of an allocation we employ the \textit{Nash Social Welfare} (NSW), a well known and regarded fairness index considered to naturally achieve a compromise between fairness and efficiency~\cite{Branzei2017} and employed in other works concerning slicing resource allocation (see~\cite{journalStoccarda} and~\cite{Caballero2017}).\newline
Here follows the definition of NSW adapted to the proposed system model:
\begin{equation}
    NSW(u_1,\dots,u_s) = \prod_{s \in \mathcal{S}} u_s^{B_s}.
\end{equation}
One can note that NSW coincides with the geometric mean of utilities weighted by provider budgets, which suggests a trade-off between overall system performances and individual fairness.\newline
It is known that EG-based formulations such as~(\referenceEG) maximize NSW. This results comes from observing that the objective function~(\ref{opt:obj}) is equivalent to the definition of NSW, formally:
\begin{equation}
    \arg\max\left\{ \sum_{s \in \mathcal{S}}B_s \log(u_s) \right\}=\arg\max\left\{ \prod_{s \in \mathcal{S}} u_s^{B_s} \right\}.
\end{equation}
As a consequence, the market equilibrium obtained through~(\referenceEG) yields the highest fairness among any feasible allocation, including solutions obtained through proportional sharing and social optimum. Additionally, Remark~\ref{remark:uniqueness} states the uniqueness of the equilibrium utilities obtained through~(\referenceEG), meaning that the proposed mechanism is capable of selecting the unique equilibrium point with the highest fairness according to NSW.\newline
This result has been summarized in the following remark:
\begin{remark}
\label{remark:max_nsw}
The utilities obtained through program~(\referenceEG) are the unique maximizer of NSW.
\end{remark}
After having characterized the NSW of the market equilibrium approach from an absolute standpoint, it is natural to question how it compares to the solution which maximizes the efficiency, i.e. the social optimum approach.\newline
On this matter, we declare the following proposition, whose proof is reported in Appendix.
\begin{proposition}
\label{proposition:unbounded_eff_loss}
The loss of NSW fairness incurred when employing the social optimum approach with respect to the market equilibrium approach is unbounded.
\end{proposition}
According to this result, it is evident how the social optimum solution, despite offering the best efficiency, presents no guarantee in terms of fairness. Indeed, Section~\ref{sec:results} shows how social optimal solutions almost always result in at least one service provider getting zero resources, which is arguably unacceptable from a quality of service standpoint.\newline
\par
In addition to a quantitative fairness analysis through NSW, we consider two well known fairness properties that characterize the qualitative fairness of the market equilibrium approach, namely \textit{Proportional Fairness} (PF) and \textit{Envy-Freeness} (EF).\newline
PF, long considered as a desirable fairness property of allocation mechanism, was first introduced in the context of bandwidth sharing among network flows~\cite{kelly1997} and has been proven to be relevant also for data-center resource sharing in~\cite{bonald2014}.\newline
A feasible utility vector $(u_1,\dots,u_S)$ is said to be proportional fair if, for any other feasible utility vector $(\bar{u_1},\dots,\bar{u_S})$, the aggregate of proportional changes is negative~\cite{MILIOTIS201617}, formally:
\begin{equation}
\label{eq:prop_fairness}
    \sum_{s \in S}B_s\frac{\bar{u_s}-u_s}{u_s}\leq 0.
\end{equation}
We characterize the proportional fairness property of the market equilibrium approach in the following proposition, whose proof is reported in Appendix.
\begin{proposition}
\label{proposition:prop_fairness}
The market equilibrium obtained through program~(\referenceEG) is proportional fair. 
\end{proposition}
\par
The final fairness property analyzed in this section is EF, which states that each agent in the market prefers its assigned bundle of resources over the bundle of any other agent~\cite{dolev2012}.\newline
This well known property owes its desirability to the stability that it implies, as agents have no reason to complain about their allocation. However, in the context of a fixed budget market model such as $\Gamma$, the discussion about EF is meaningful only when service providers all have the same budget.\newline
Under this assumption, resource bundles $\left(\mathbf{X}_1,\dots,\mathbf{X}_S\right)$ and $\left(\mathbf{Y}_1,\dots,\mathbf{Y}_S\right)$ are considered envy-free if $u_s(\mathbf{X}_s,\mathbf{Y}_s)\geq u_s(\mathbf{X}_t,\mathbf{Y}_t)$ for any $s,t \in \mathcal{S}$.\newline
The proposition that follows claims the envy-freeness property of the market equilibrium.
\begin{proposition}
\label{proposition:envy_freeness}
The ME obtained through~(\referenceEG) when service providers all have equal budget is envy-free. 
\end{proposition}
The proof is straightforward once one notes that equilibrium allocations are mutually affordable, given that all providers experience the same resource prices and have equal budgets. Hence each provider can afford to buy any opponent's resource bundle, but still prefers its own since it is the one that maximizes utility. 

%% file: Content/Results/results.tex
In this section we numerically evaluate the performance and fairness of the allocation approaches proposed in section~\ref{sec:alloc_approaches} applied to different instances of the system model. Such instances are characterized by parameters that vary according to the specific result or property that is intended to be highlighted, however some assumptions will be valid throughout the entire section, unless otherwise stated.\newline 
In particular, we consider ENs to offer CPU and RAM capacity and, as such, $D_{m,r}^\text{MEC}$ is measured in number of CPUs or RAM size and service requirements $d_{s,r}^\text{MEC}$ in the MEC domain are likewise measured, namely number of CPU cores required for timely execution and RAM occupation.\newline
Network resources are considered as portions of the total bandwidth available in a specific cell and parameters $d_{s,c}^\text{RAN}$ and $D_c^\text{RAN}$ are numerically represented in Hertz (Hz).\newline
Simulation instances are generated in MATLAB and the proportional sharing allocation mechanism was implemented in the same environment, while Market Equilibrium, Social Optimum and Weighted Social Optimum mechanisms are executed employing solvers IPOPT~\cite{Wachter2006} and CPLEX, respectively.
\begin{figure*}[!t]
    \centering
    \subfloat[Cumulative distribution function of provider utility.]{\includegraphics[width=2.2in]{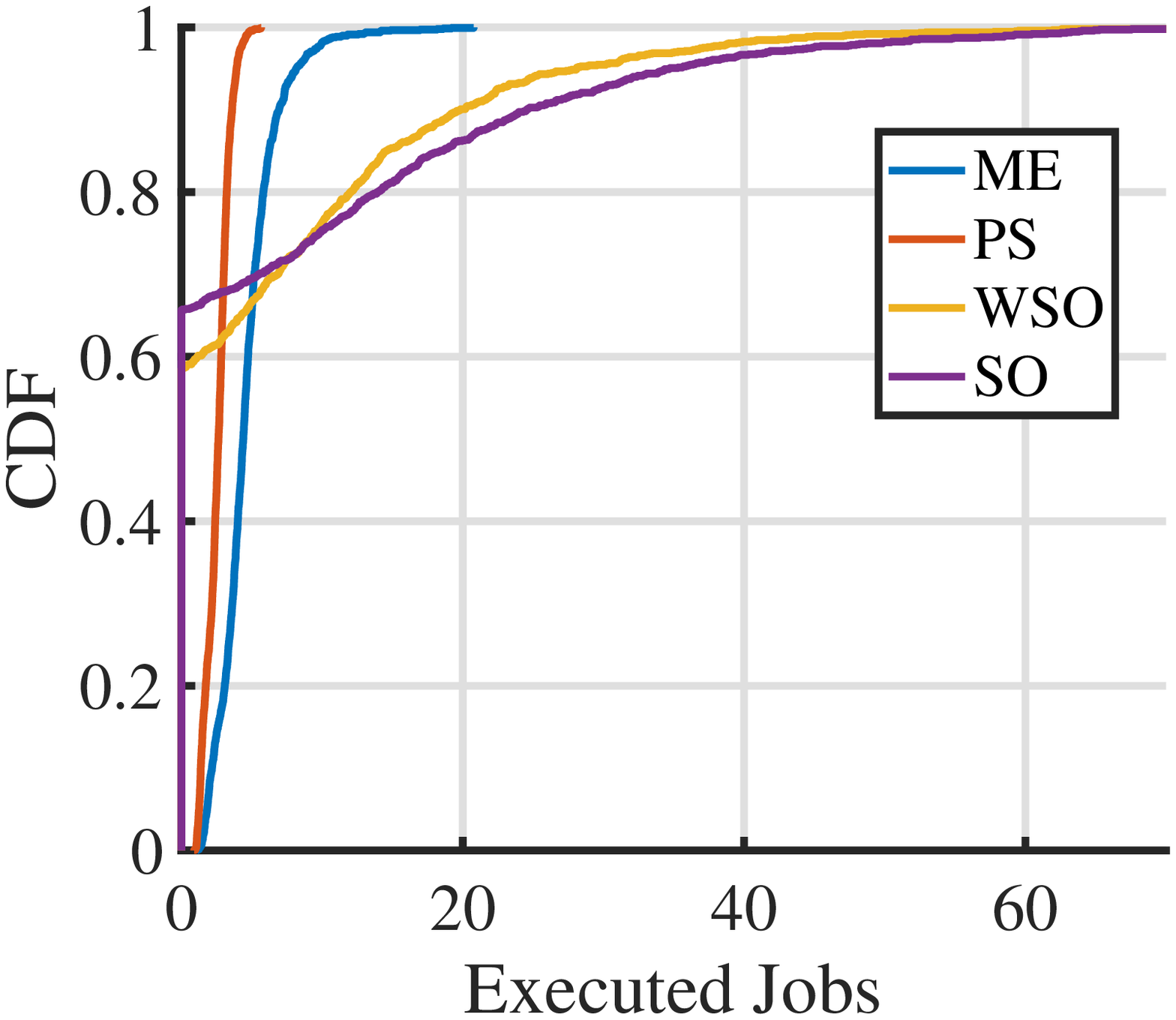}%
\label{fig:cdf_per_service}}
\hfil
\subfloat[Cumulative distribution function of concurrently executed jobs in the system.]{\includegraphics[width=2.2in]{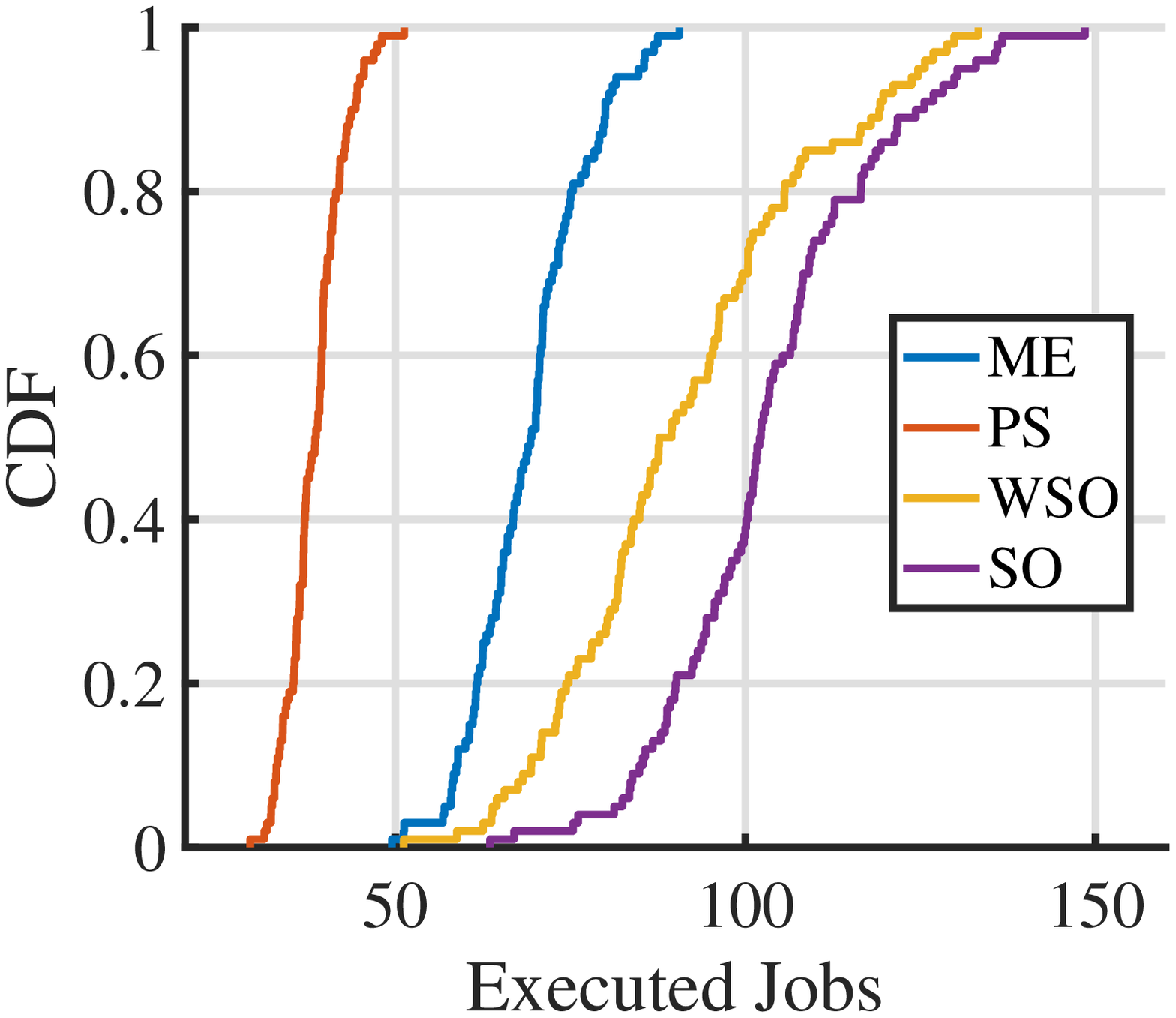}%
\label{fig:cdf_tot}}
\hfil
\subfloat[Cumulative distribution function of ME and PS efficiency]{\includegraphics[width=2.2in]{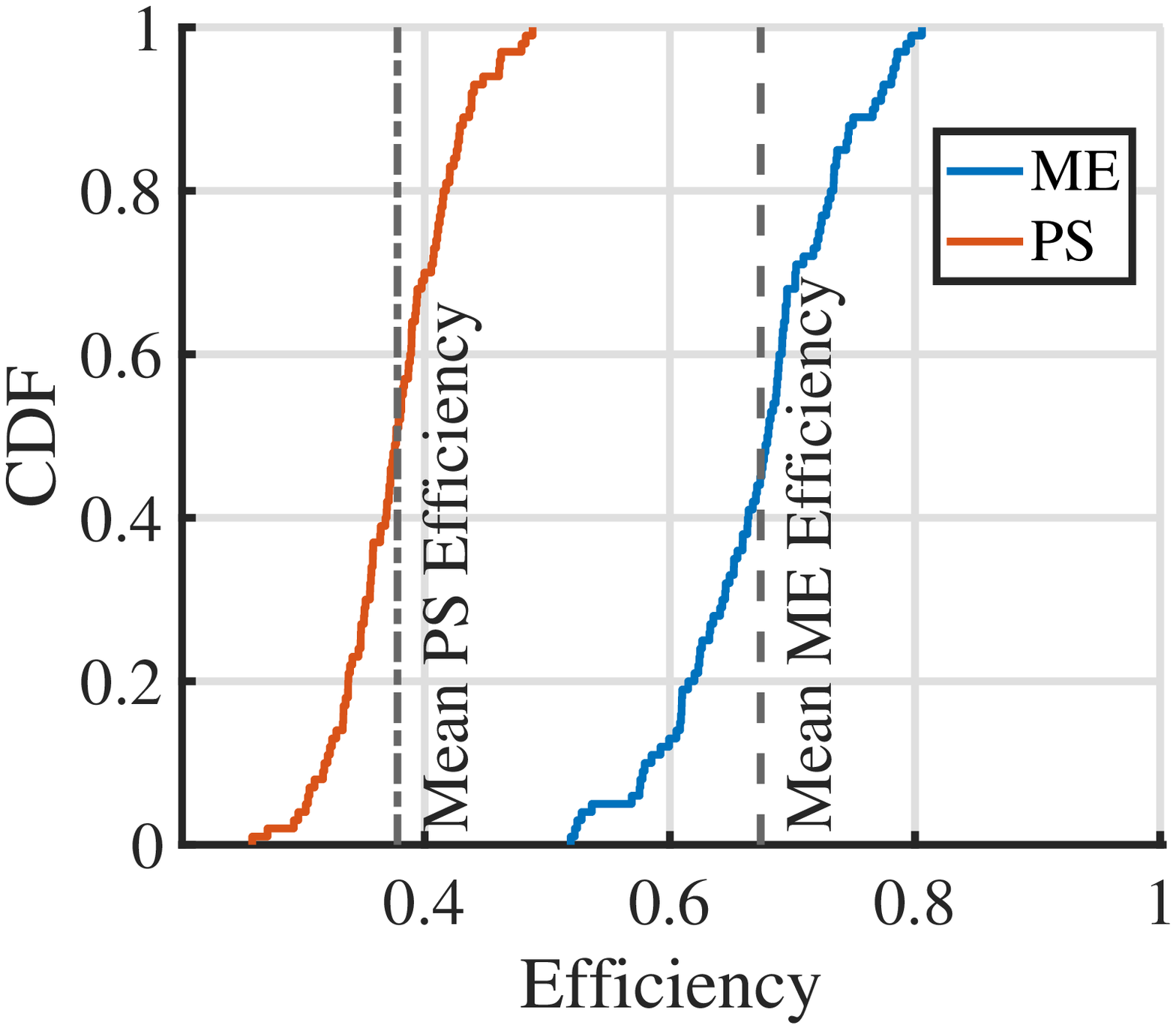}%
\label{fig:cdf_eff}}
\caption{Cumulative distribution functions of some performance parameters.}
\label{fig:cdfs}
\end{figure*}
\subsection{System Performance and Fairness}
We consider a localized MEC/RAN system comprising of heterogeneous network cells and computation nodes.\newline
In the RAN domain we identify 2 large cells with 40MHz of ccapacity and 5 smaller cells with 20MHz of capacity.\newline
In the MEC domain we identify 2 types of EN, whose computation resources are either 32 CPU cores and 128Gb of RAM, for what we call a \textit{CPU node}, or 16 CPU cores and 256Gb of RAM, for what we call \textit{RAM node}. Our simulated system includes 5 \textit{CPU nodes} and 5 \textit{RAM nodes}.\newline
In order to capture the heterogeneity of resource requirements of different services, we define 4 service templates that represent possible demands configurations in a cloud computing scenario. In details, we define 3 templates representing CPU-intensive, RAM-intensive and bandwidth-intensive services and an additional more balanced service with overall high resource requirements. Numerical values of service templates are detailed in Table~\ref{tab:templates}.
\begin{table}[t!]
\centering
\caption{Service Templates}
\label{tab:templates}
\begin{tabular}{|l|l|l|l|l|} 
\hline
Name          & $d_{s,\text{CPU}}^\text{MEC}$ & $d_{s,\text{RAM}}^\text{MEC}$ & $d_{s,c}^\text{RAN}$ & $B_s$  \\ 
\hline
CPU-Intensive & 4 CPUs      & 8Gb         & 3MHz            & 1      \\ 
\hline
RAM-Intensive & 1 CPU       & 32Gb         & 3MHz            & 1      \\ 
\hline
BW-Intensive  & 1 CPU       & 8Gb         & 10MHz           & 1.5    \\ 
\hline
Balanced      & 5 CPUs      & 40Gb         & 5MHz            & 2      \\
\hline
\end{tabular}
\end{table}
\newline
In each simulation run, one among the aforementioned templates is given at random to each of the 15 service providers competing for system resources. Furthermore, Gaussian noise is added to each resource requirements with mean 0 and variance equal to $25\%$ of the original numerical value. This has been done to account for requirements oscillations given by unpredictable variations in payload size and computational difficulty, as well as in network conditions.\newline
\par
Figure~\ref{fig:cdfs} shows the cumulative distribution functions of some performance indicators extracted from 100 simulated instances. In particular, Figure~\ref{fig:cdf_per_service} allows to immediately appreciate the impact of different allocation approaches on the number of concurrent jobs of each single service provider, i.e. its utility.\newline
Social optimum-based solutions present zero executed jobs (i.e. zero allocated resources) for at least 60\% of the cases, while the rest of the providers obtain a relative high performance. This behaviour is indeed expected in this type of approach, since the solution tends to favor those provider who can simultaneously execute the most jobs with less resources.\newline
On the other hand, ME and PS approaches always guarantee that each SP can execute some jobs, even if this means a lower maximum per-provider performance. Furthermore, this particular figure shows how providers always get better performance with the ME approach in respect of PS, confirming the \textit{sharing incentive} property implied by Theorem~\ref{theo:shar_inc}.\newline
Figure~\ref{fig:cdf_tot} shows the cumulative distribution functions of sum of per-provider concurrently executable jobs, namely the system performance, for each of the studied allocation approaches.\newline
As expected, SO outperforms the other approaches and ME consistently delivers better performance than PS, which is shown to be the least performing. This is also evident in Figure~\ref{fig:cdf_eff}, where the distribution of efficiencies $\eta_{ME}$ and $\eta_{PS}$ are plotted. Here it is shown how the ME approach is about $30\%$ more efficient than PS in average, a result that again confirms the efficiency remarks of Section~\ref{sec:market_mod_props}. Furthermore, ME presents a worst-case efficiency of $52\%$, higher than the theoretical lower bound of $26\%$ given by Remark~\ref{remark:efficiency_bound}.\newline
\begin{figure}[t]
    \centering
    \includegraphics[width=2.4in]{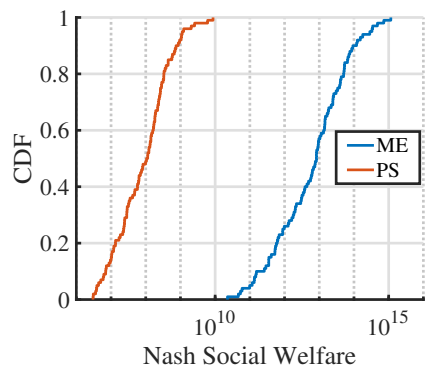}
\caption{Nash Social Welfare cumulative distribution function of ME and PS.}
\label{fig:cdf_nsw}
\end{figure}
\par
Figure~\ref{fig:cdf_nsw} shows the cumulative distribution function of the NSW for both ME and PS approaches. Plots corresponding to social optimum-based approaches are missing because such solutions almost always present at least one service provider with 0 concurrently executed jobs, yielding NSW values that are mainly 0.\newline
Moreover, the horizontal axis of this figure is in logarithmic scale due to both a large difference between worst-case and best-case values and the gap between the two distributions. This behaviour is to be expected, since the considered fairness index exponentially increases with both budgets and number of service providers.
Additionally, Figure~\ref{fig:cdf_nsw} confirms Remark~\ref{remark:max_nsw} by showing how NSW of market equilibrium approach is much larger than the proportional sharing case in all simulated instances.
\subsection{Sensitivity Analysis}
\begin{figure}[!t]
\centering
\subfloat[ME single service utility vs budget ]{\includegraphics[width=1.7in]{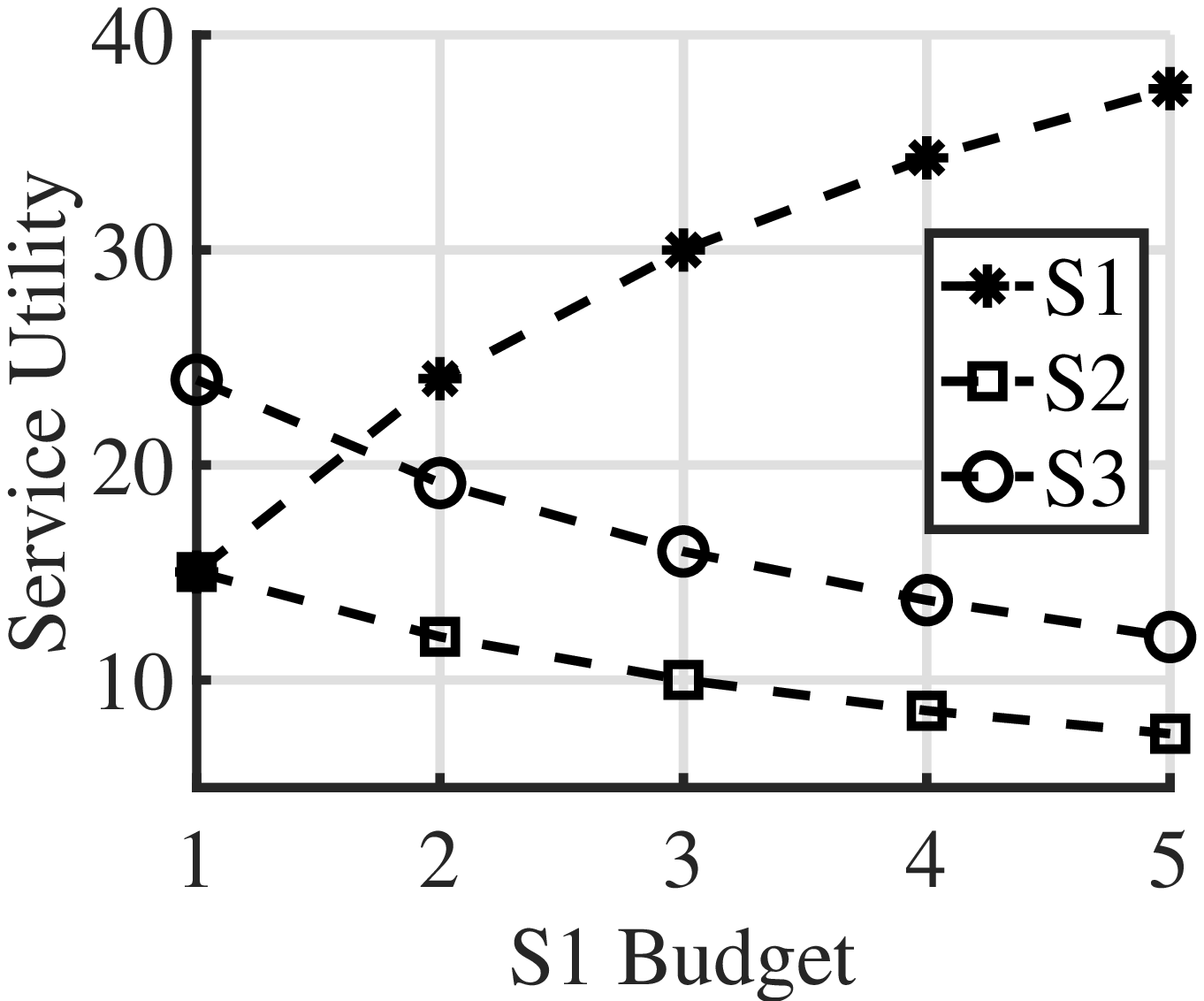}%
\label{fig:service_utility_variation_budget}}
\hfil
\subfloat[System performance vs budget]{\includegraphics[width=1.7in]{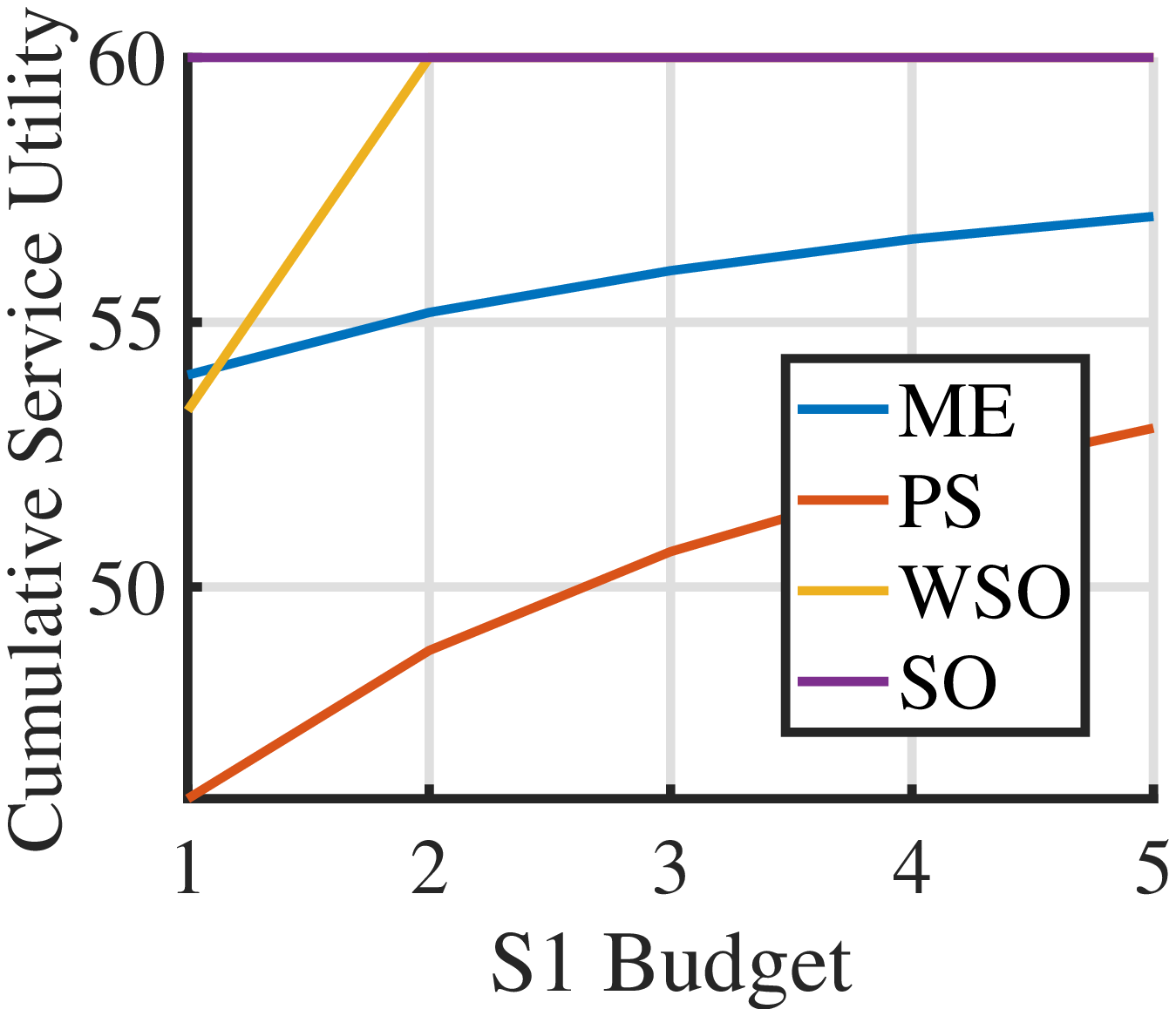}%
\label{fig:sys_perf_variation_budget}}
\hfil
\subfloat[ME single service utility vs $M$ ]{\includegraphics[width=1.7in]{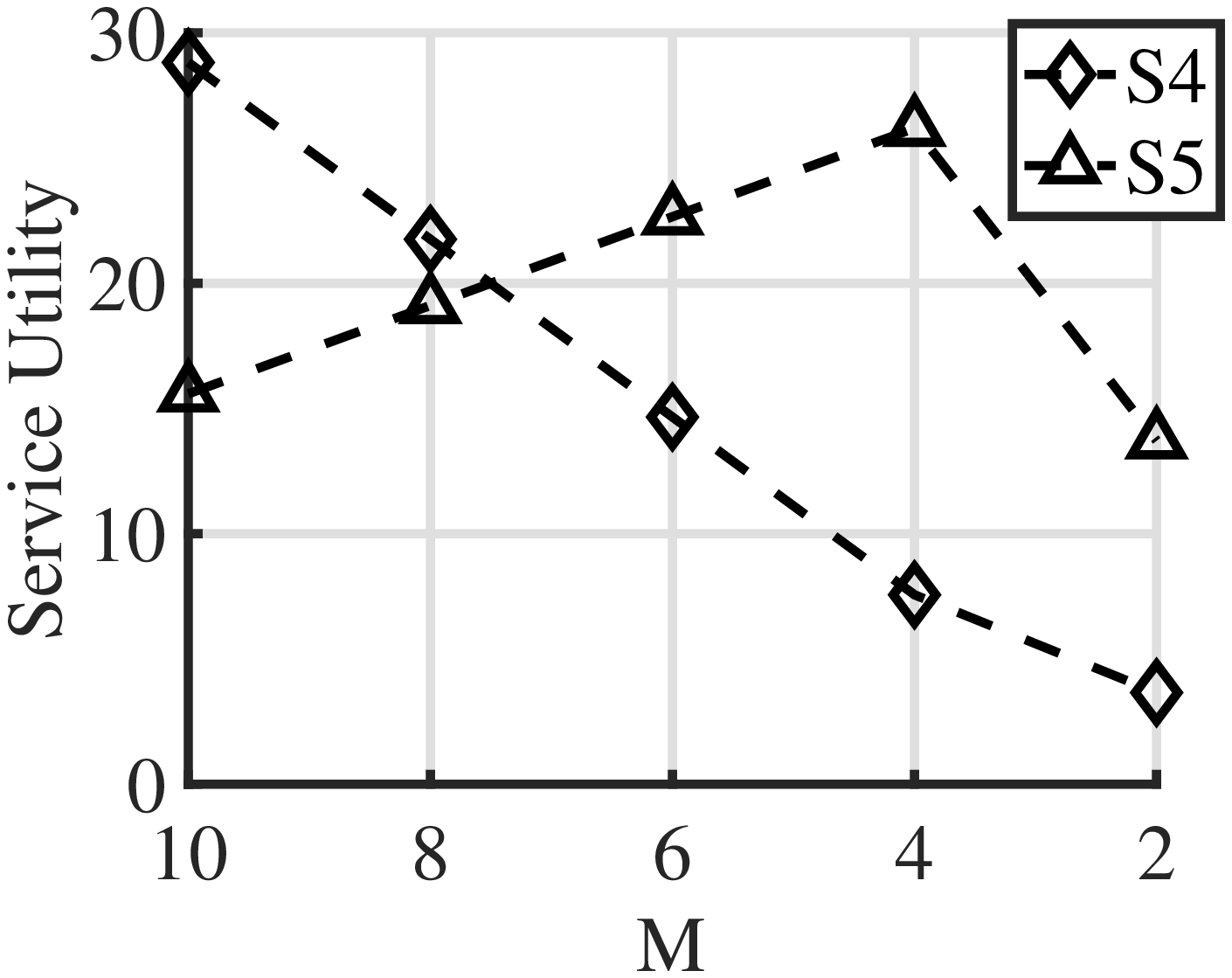}%
\label{fig:m_variations}}
\hfil
\subfloat[ME single service utility vs $C$]{\includegraphics[width=1.7in]{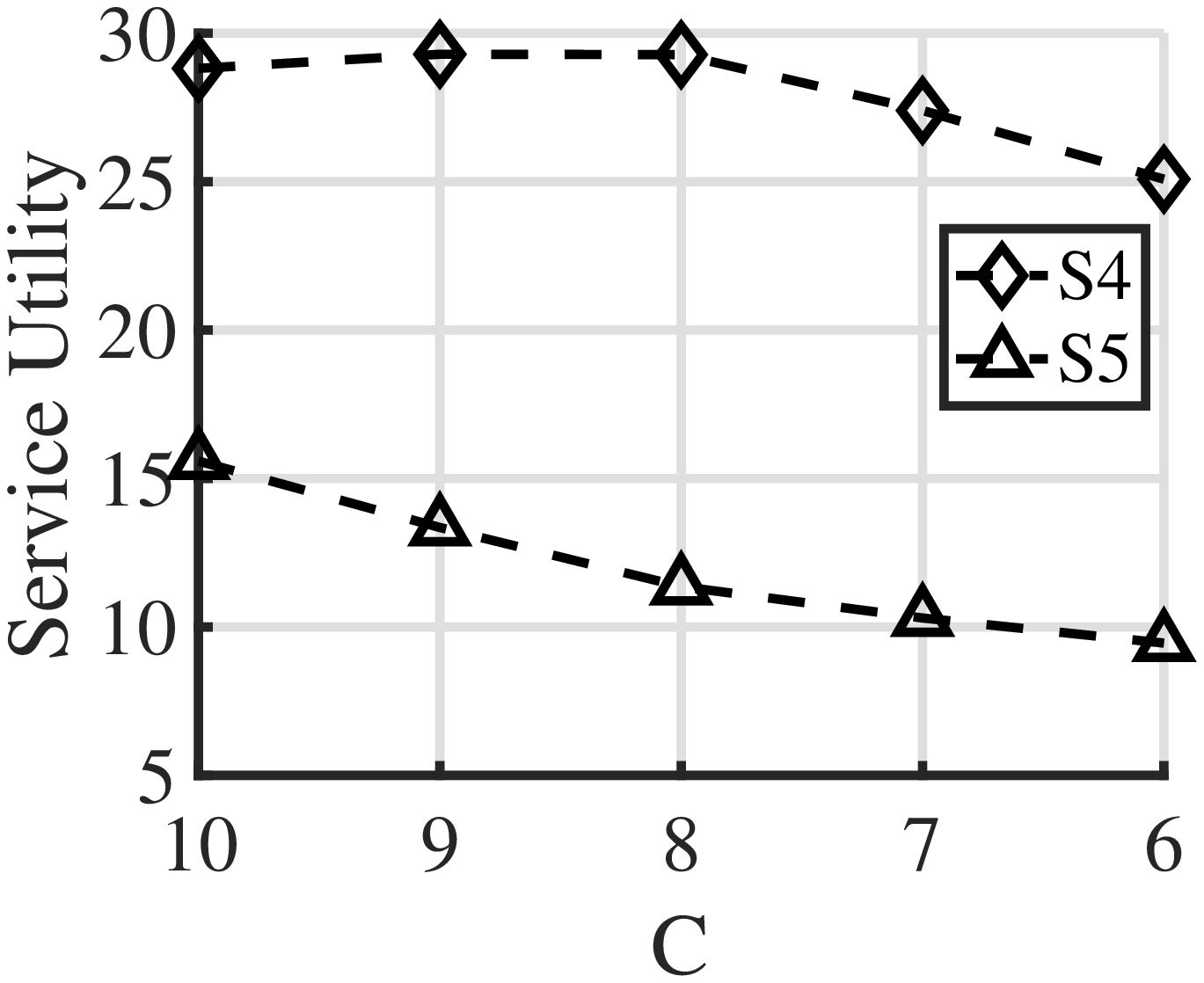}%
\label{fig:c_variations}}
\caption{Sensitivity Analysis}
\label{fig:sens_analisys}
\end{figure}

We start by analysing the sensitivity of the different allocation mechanisms to variations in budget values from the per-provider performance standpoint and overall system performance.\newline
Consider the previously detailed system deployment and 3 provider associated with different services: 2 CPU-intensive services and one Balanced service, which will be called \textit{S1}, \textit{S2}, and \textit{S3} respectively.\newline
Figures~\ref{fig:service_utility_variation_budget} and~\ref{fig:sys_perf_variation_budget} show the variations in performance when the budget of \textit{S1} is allowed to increase from its default value of 1 up to 5.\newline
In particular, Figure~\ref{fig:service_utility_variation_budget} demonstrates how single SP utilities adapt to the budget variations in the ME mechanism. When \textit{S1} and \textit{S2} have equal budgets, they obtain the same utility since their requirements are symmetrical and the mechanism cannot prefer either of the two. As \textit{S1} budget increases, the allocation mechanism grants her increasingly more resources, confirming the service prioritization effect of the budget values. When \textit{S1} and \textit{S3} budgets are the same, i.e. 2, the mechanism demonstrates efficiency by granting more resources to \textit{S1}, which can execute more jobs under the same allocation conditions because it requires less per-job resources.\newline
Figure~\ref{fig:sys_perf_variation_budget} shows the impact of \textit{S1} budget variations on the cumulative provider utility when employing different allocation mechanisms. As expected, SO remains unaffected, while the other mechanisms show an increment in overall performance due to a more efficient resource utilization when increasingly larger allocations are granted to \textit{S1}.\newline
\par
Next, we present an analysis of the service provider utilities obtained through ME when the number of nodes and cells in the system is allowed to vary. In this case, consider a system comprising 5 \textit{small cells} and 5 \textit{large cells} and the same MEC deployment as the previous analyses. In this scenario, one Balanced service and one BW-intensive service populate the system, offered by service providers \textit{S4} and \textit{S5} respectively.\newline
Leaving the RAN domain unchanged, we allow for $M$ to decrease from $10$ to $2$ by iteratively excluding one \textit{CPU node} and one \textit{RAM node} at the same time.\newline
Figure~\ref{fig:m_variations} shows the impact on the utility of the two considered service providers. As the computational capacity of the system starts decreasing, \textit{S4} sees a decrease in the executed jobs which in turns frees some radio access capacity that can be utilized by bandwidth-hungry \textit{S5}, giving her a performance boost. This behavior continues until the computational capacity is so low that it forms a bottleneck for both providers, showing in practice the presence of the domain bottleneck phenomenon that has been introduced in Section~\ref{sec:sys_model}.\newline
Coming back to the original deployment, we analyze the impact of RAN capacity variations by decreasing the number of \textit{small cells} from 5 to 1, while leaving the rest untouched.\newline
Figure~\ref{fig:c_variations} illustrates a behaviour similar to the previous analysis, even if less accentuated. Indeed, as the radio resources decrease, \textit{S5} hits a resource bottleneck and her performance decrease. This causes more computational resources to be available to \textit{S4}, allowing for a small but noticeable performance increase. As seen before, the domain bottleneck is eventually hit and both service providers start experiencing lower performance.

%% file: Content/Conclusion/conclusion.tex
In this work we have analyzed the problem of efficient and fair joint management of radio and computing resource in a MEC/RAN deployment. On this matter, we have proposed a techno-economic market model where service providers sharing the same physical infrastructure can purchase resources, while being constrained to a budget limit. 
We formulated a convex program that computes the market equilibrium for any instance of the market model and proceeded to extensively analyze the properties of such solution in terms of efficiency and fairness, showing in particular how the properties of \textit{Pareto Optimality}, \textit{Nash Social Welfare} maximization, \textit{Proportional Fairness}, \textit{Sharing Incentive} and \textit{Envy-Freeness} apply. As it is usual for game theory-based works, we also give a characterization of the price of anarchy of the solution.
Additionally, we propose alternative multi-resource allocation mechanisms which are not based on any market concept and we compare them with the market equilibrium approach.
Finally, we present numerous results coming from a simulation of the proposed allocation mechanisms, in order to numerically analyze and compare the performance of the mechanisms, confirm the theoretical properties of the market model and give an insight on the sensitivity of the solution to parameters variations.

%% file: Content/Appendices/appendix.tex
\subsection{Theorem~\ref{theo:me_solution}}
\input{Content/Appendices/dual}
\par
\textit{Proof of Theorem~\ref{theo:me_solution}.}
\input{Content/Appendices/theo1}
\qed
\subsection{Theorem~\ref{theo:shar_inc}}
\textit{Proof of Theorem~\ref{theo:shar_inc}.}
\input{Content/Appendices/theo2}

\qed
\subsection{Propositions~\ref{proposition:unbounded_eff_loss} and~\ref{proposition:prop_fairness}}
\par
\textit{Proof of Proposition~\ref{proposition:unbounded_eff_loss}.}
\input{Content/Appendices/propos1}
\qed\newline
\par
\textit{Proof of Proposition~\ref{proposition:prop_fairness}.}
\input{Content/Appendices/propos2}
\qed
%

%% file: Content/Appendices/dual.tex
Before giving the proof, we write the Lagrangian dual of~(\referenceEG):
\begin{align}
    L(x,y,u,j,p,\mathbf{\lambda},\nu)=\sum_{s \in \mathcal{S}}B_s \log(u_s)\\
    +\sum_{m,r}p_{m,r}^\text{MEC}(D_{m,r}^\text{MEC}-\sum_{s}x_{s,m,r})\\
    +\sum_c p_c^\text{RAN}(D_c^\text{RAN}-\sum_{s}y_{s,c})\\
    +\sum_{s,m,r} \lambda_{s,m,r}^b(\frac{x_{s,m,r}}{d_{s,r}^\text{MEC}} - j_{s,m})\\
    +\sum_s \lambda_s^d(\sum_{m}j_{s,m} - u_s)\\
    +\sum_s \lambda_s^e(\sum_{c}\frac{y_{s,c}}{d_{s,c}^\text{RAN}}- u_s)\\
    +\sum_{s,m,r} \nu^f_{s,m,r}x_{s,m,r}+\sum_{s,c}\nu^g_{s,c}y_{s,c}+\sum_{s,m}\nu^h_{s,m}j_{s,m}.
\end{align}
If system capacities $D_{m,r}^\text{MEC}$ and $D_c^\text{RAN}$ are strictly positive, then it is possible to find an interior point of~(\referenceEG) for which all the inequalities are strictly satisfied. In this case, Slater's condition can be applied and strong duality holds, thus KKT conditions are to be satisfied for any optimal solution~\cite{boyd2004convex}. We write some of these that will be instrumental for proving the results in this appendix. In particular, we 
consider some stationarity conditions:
\begin{flalign}
&\frac{B_s}{u_s} -(\lambda_s^d+ \lambda_s^e) = 0,&\forall s \in \mathcal{S} \label{ktt:1}\\
&\frac{\lambda^b_{s,m,r}}{d_{s,r}^\text{MEC}}-p_{m,r}^\text{MEC} + \nu^f_{s,m,r} = 0,&\forall s \in \mathcal{S}, m \in \mathcal{M}, r \in \mathcal{R} \label{ktt:2}\\
&\frac{\lambda_s^e}{d_{s,c}^\text{RAN}} - p_c^\text{RAN} +\nu^g_{s,c}= 0,&\forall s \in \mathcal{S}, c \in \mathcal{C} \label{ktt:3}\\
&\lambda_s^d -\sum_r \lambda_{s,m,r}^b + \nu^h_{s,m}= 0,&\forall s \in \mathcal{S}, m \in \mathcal{M},\label{ktt:4}
\end{flalign}
some complementarity slackness conditions:
\begin{flalign}
&p_{m,r}^\text{MEC}(D_{m,r}^\text{MEC}-\sum_{s}x_{s,m,r}) = 0,&\forall m \in \mathcal{M}, \forall r \in \mathcal{R} \label{ktt:5}\\
&p_c^\text{RAN}(D_c^\text{RAN}-\sum_{s}y_{s,c}) = 0,&\forall c \in \mathcal{C},\label{ktt:6}\\
&x_{s,m,r}\nu_{s,m,r}^f = 0,&\forall s \in \mathcal{S}, m \in \mathcal{M}, r \in \mathcal{R},\label{ktt:new1}\\
&y_{s,c}\nu_{s,c}^g = 0,&\forall s \in \mathcal{S}, c \in \mathcal{C},\label{ktt:new2}\\
&j_{s,m}\nu_{s,m}^h = 0,&\forall s \in \mathcal{S}, m \in \mathcal{M},\label{ktt:new3}
\end{flalign}
and finally some for dual feasibility:
\begin{flalign}
\label{ktt:7}
    &\nu_{s,m,r}^f,\, \nu_{s,c}^g,\, \nu_{s,m}^h \geq 0,&\forall s \in \mathcal{S}, m \in \mathcal{M}, r \in \mathcal{R}, c \in \mathcal{C}.
\end{flalign}

%% file: Content/Appendices/theo1.tex
It is sufficient to show that market equilibrium conditions reported in section~\ref{sec:alloc_approaches} are satisfied by the optimal solutions of~(\referenceEG). We start by showing that the optimal resource allocations maximize the \textit{bang-per-buck} of each provider, i.e. SPs spend budget where they obtain the highest marginal utility increment. We start by rewriting condition~(\ref{ktt:1}) as follows:
\begin{flalign}
    &\frac{B_s}{u_s}=\lambda_s^d+\lambda_s^e,&\forall s \in \mathcal{S}. \label{me_proof:1}
\end{flalign}
From~(\ref{ktt:3}) and~(\ref{ktt:7}) we get:
\begin{flalign}
    &\lambda_s^e \leq p_c^\text{RAN} d_{s,c}^\text{RAN},&\forall s \in \mathcal{S}, c \in \mathcal{C},\label{me_proof:3}
\end{flalign}
which holds with equality for $y_{s,c}>0$ due to~(\ref{ktt:new2}).\newline
Similarly, from~(\ref{ktt:4}) and~(\ref{ktt:7}) we get:
\begin{flalign}
    &\lambda_s^d\leq \sum_r\lambda_{s,m,r}^b,&\forall s \in \mathcal{S}, m \in \mathcal{M},
\end{flalign}
that again holds with equality if $j_{s,m}>0$ due to~(\ref{ktt:new3}).\newline
From~(\ref{ktt:2}) and~(\ref{ktt:7}) we get:
\begin{flalign}
    &\lambda^b_{s,m,r}\leq p_{m,r}^\text{MEC}d_{s,r}^\text{MEC},&\forall s \in \mathcal{S}, m \in \mathcal{M}, r \in \mathcal{R}\label{me_proof:2}
\end{flalign}
holding with equality if $x_{s,m,r}>0$ due to~(\ref{ktt:new1}).\newline
Substituting in~(\ref{me_proof:1}) we get:
\begin{flalign}
    &\frac{B_s}{u_s}\leq \sum_r p_{m,r}^\text{MEC}d_{s,r}^\text{MEC}+p_c^\text{RAN} d_{s,c}^\text{RAN},&\forall s \in \mathcal{S}, m \in \mathcal{M}, c \in \mathcal{C}. \label{me_proof:4}
\end{flalign}
Let $q_{s,m}=\sum_r p_{m,r}^\text{MEC}d_{s,r}^\text{MEC}$, this corresponds to the cost of scheduling one job in MEC node $m$ for service $s$. Note that the inverse of this quantity is the bang-per-buck of MEC server $m$ for service provider $s$, since it is the ratio between an unitary increment in MEC-processed jobs and the price of scheduling said job.\newline
Similarly, $w_{s,c}=p_c^\text{RAN} d_{s,c}^\text{RAN}$ is the cost of uploading one job through cell $c$ for service $s$. Again, the inverse of $w_{s,c}$ is the bang-per-buck in cell $c$ for provider $s$. It is now clear that minimizing $q_{s,m} + w_{s,c}$ for any node-cell couple means maximizing the \textit{bang-per-buck} of that couple. Therefore, we define $q_{min}$ as the minimum cost of executing one job over any node-cell couple in the entire system.\newline
Finally from~(\ref{me_proof:4}) we get:
\begin{flalign}
    &\frac{B_s}{u_s} \leq q_{s,m} + w_{s,c},&\forall s \in \mathcal{S}, m \in \mathcal{M}, c \in \mathcal{C}\label{eq:prices_bound}
\end{flalign}
suggesting that $\frac{B_s}{u_s} \leq q_{min}$.\newline
We conclude this part of the proof noting that for each variable $x_{s,m,r}>0, y_s^c>0, j_{s,m}>0$, Eq.~(\ref{eq:prices_bound}) holds with equality, meaning that each service provider buys only those resources yielding the best \textit{bang-per-buck}.\newline
Now for the second condition, we verify that resources are either priced or not fully allocated by noting that dual variables $p_{m,r}^\text{MEC}$ and $p_c^\text{RAN}$ are either positive if the corresponding resource is allocated to its capacity, or they are zero, as shown by the complementary slackness conditions~(\ref{ktt:5}) and~(\ref{ktt:6}).\newline
Finally, for a given optimal solution, only strictly positive variables contribute to the budget expenses. This means that we can rewrite~(\ref{eq:prices_bound}) as follows:
\begin{flalign}
        &\frac{B_s}{u_s} = q_{s,m} + w_{s,c},&\forall s \in \mathcal{S}, m \in \mathcal{M}, r \in \mathcal{R}. 
\end{flalign}
We note that the price of executing one job $q_{s,m} + w_{s,c}$ is constant with respect to the cell and node employed for its execution. Expressing this quantity as $q_s$ allows us to rewrite the previous equation:
\begin{flalign}
    &B_s = u_s q_s,&\forall s \in \mathcal{S},
\end{flalign}
meaning that the budget is fully allocated when the solution is optimal.

%% file: Content/Appendices/theo2.tex
It is sufficient to show that proportional sharing allocations are both feasible in terms of capacity and exhaust the provider budgets given the equilibrium prices. In this case, such allocations would be chosen as solution of~(\referenceEG) if they were to maximize utilities, thus $u_s^*\geq \hat{u}_s$ for each $s \in\ \mathcal{S}$\newline
The first claim is trivial, we proceed to prove the second.\newline
Let $x_{s,m,r},\,y_{s,c}$ and $p_{m,r}^\text{MEC},\, p_{c}^\text{RAN}$ be the market equilibrium allocations and prices, respectively. Then, given the budget exhaustion property, it must be that:
\begin{equation}
\label{eq:th2proof1}
    \sum_s B_s = \sum_{s,m,r} x_{s,m,r}p_{m,r}^\text{MEC} + \sum_{s,c} y_{s,c}p_c^\text{RAN}.
\end{equation}
Due to the market clearing condition, prices are such that either $\sum_{s}x_{s,m,r}=D_{m,r}^\text{MEC}$ or $p_{m,r}^\text{MEC}=0$ and either $\sum_s y_{s,c} = D_c^\text{RAN}$ or $p_c^\text{RAN}=0$, allowing us to rewrite~(\ref{eq:th2proof1}) as follows:
\begin{equation}
\sum_s B_s=\sum_{m,r} D_{m,r}^\text{MEC}p_{m,r}^\text{MEC} + \sum_{c} D_{c}^\text{RAN}p_c^\text{RAN}.
\end{equation}
Now multiplying both sides by $\frac{B_s}{\sum_s B_s}$ we obtain:
\begin{equation}
B_s =\sum_{m,r}\frac{B_s}{\sum_s B_s} D_{m,r}^\text{MEC}p_{m,r}^\text{MEC} + \sum_{c} \frac{B_s}{\sum_s B_s}D_{c}^\text{RAN}p_c^\text{RAN},
\end{equation}
meaning that proportional sharing allocations exhaust the budget of each service.

%% file: Content/Appendices/propos1.tex
Consider an instance in which all services have equal resource requirements and budgets. In this case, a solution where all the capacity of the system is reserved for one service only would be acceptable as social optimum and the corresponding NSW would be $0$.\newline
On the other hand, a proportional sharing solution would be equally optimal and, by Theorems~\ref{theo:me_solution} and~\ref{theo:shar_inc}, the proportional sharing utilities would coincide with the utilities of the market equilibrium approach.\newline
For these utilities, NSW would be strictly positive and since it is increasing in $B_s$ by definition, the fairness loss can be increased arbitrarily for this instance.

%% file: Content/Appendices/propos2.tex
Note that equation~(\ref{eq:prop_fairness}) can be rewritten as:
\begin{equation}
    \sum_{s \in \mathcal{S}} B_s \frac{d}{du_s}\left[ \log(u_s) \right]\cdot du_s \leq 0,
\end{equation}
which is true only if $(u_1,\dots,u_s)$ maximizes $\sum_{s \in \mathcal{S}}B_s\log(u_s)$, hence the proof.

%% file: Content/Authors Bio/authors_bio.tex
%

\begin{IEEEbiography}[{\includegraphics[width=1in,height=1.25in,clip,keepaspectratio]{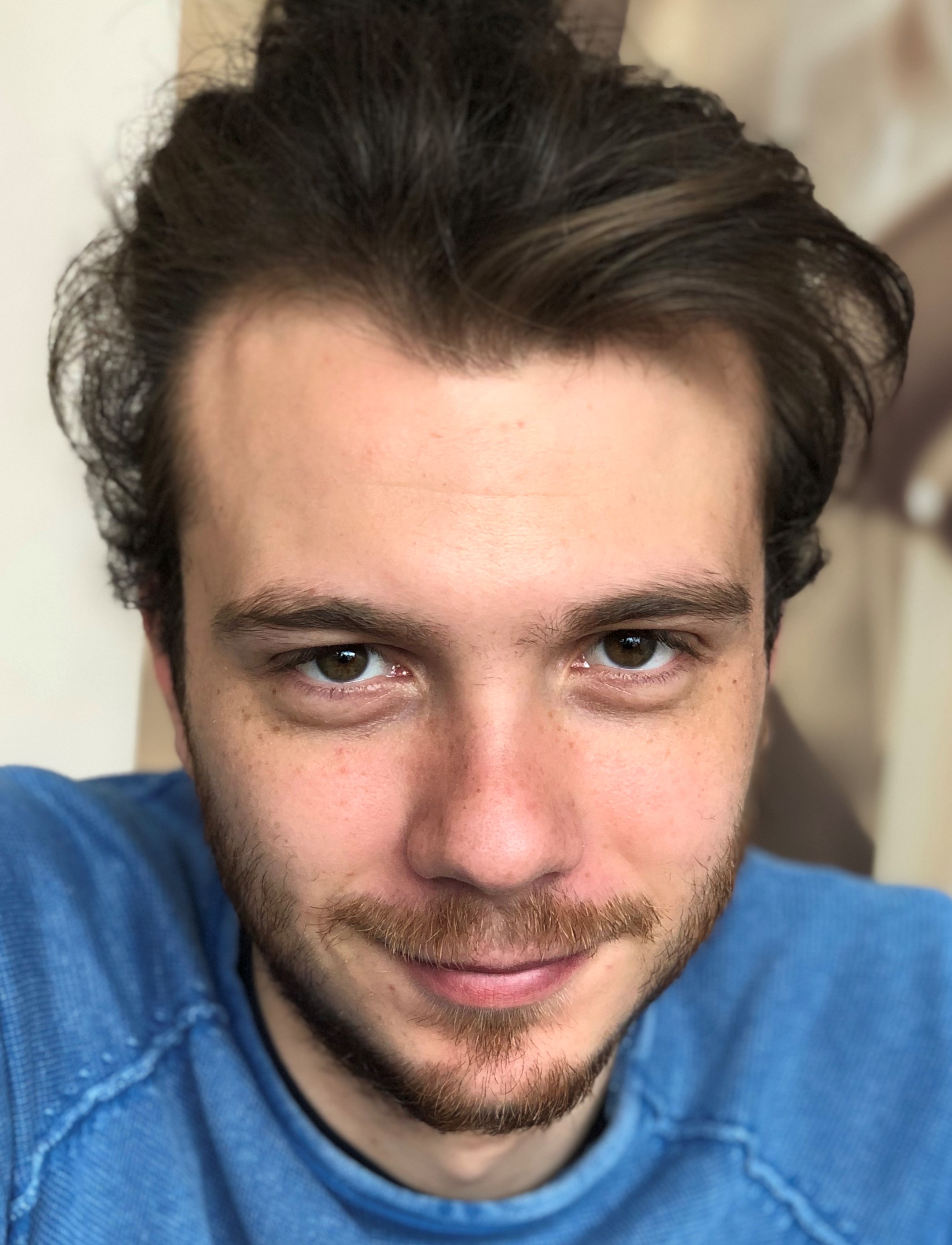}}]{Eugenio Moro}
is a PhD student at Politecnico di Milano, Department of Electronics, Information and Bioengineering. His research area is Telecommunication, with a focus on optimization techniques and game theory applied to wireless networks. In 2016, he received his bachelor’s degree in information engineering at Università del Salento. In 2017, he was enrolled in the MSc course of Telecommunications Engineering at Politecnico di Milano, where he graduated in 2019.
\end{IEEEbiography}

\begin{IEEEbiography}[{\includegraphics[width=1in,height=1.25in,clip,keepaspectratio]{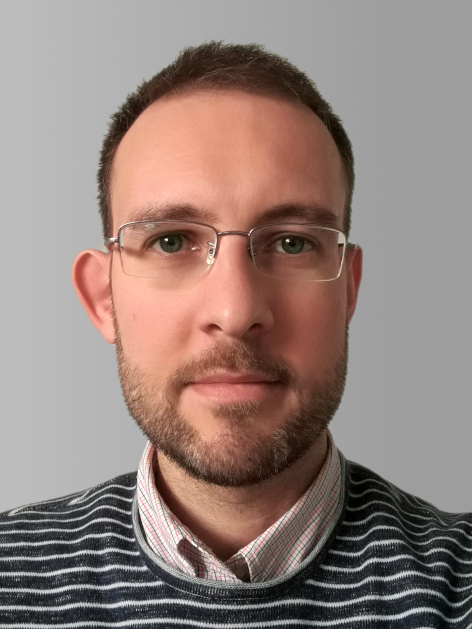}}]{Ilario Filippini} (S’06–M’10–SM’16) received B.S. and M.S. degrees in Telecommunication Engineering and a Ph.D in Information Engineering from the Politecnico di Milano, in 2003, 2005, and 2009, respectively. He is currently an Associate Professor with the Dipartimento di Elettronica, Informazione e Bioingegneria, Politecnico di Milano. His research interests include planning, optimization, and game theoretical approaches applied to wired and wireless networks, performance evaluation and resource management in wireless access networks, and traffic management in software defined networks. On these topics, he has published over 60 peer-reviewed articles. He serves in the TPC of major conferences in networking and as an Associate Editor of \textit{IEEE Transactions on Mobile Computing} and \textit{Elsevier Computer Networks}.
\end{IEEEbiography}



\vfill
